# Brownmillerite $Ca_2Co_2O_5$: Synthesis, Stability, and Re-entrant Single-Crystal-to-Single-Crystal Structural Transitions

Junjie Zhang,[*,†] Hong Zheng,[†] Christos D. Malliakas,[†] Jared M. Allred,[†] Yang Ren,[§] Qing'an Li,[†] Tian-Heng Han[†,‡] and J.F. Mitchell[†]

[†]Materials Science Division, Argonne National Laboratory, Argonne, Illinois 60439, USA
[§]X-ray Science Division, Argonne National Laboratory, Argonne, Illinois, 60439, USA
[‡]The James Franck Institute and Department of Physics, University of Chicago, Chicago, IL 60637, USA

**ABSTRACT**

$Ca_2Co_2O_5$ in the brownmillerite form was synthesized using a high-pressure optical-image floating zone furnace, and single crystals with dimensions up to $1.4 \times 0.8 \times 0.5$ mm$^3$ were obtained. At room temperature, $Ca_2Co_2O_5$ crystallizes as a fully ordered brownmillerite variant in the orthorhombic space group *Pcmb* (No. 57) with unit cell parameters $a$=5.28960(10) Å, $b$=14.9240(2) Å, and $c$=10.9547(2) Å. With decreasing temperature, it undergoes a re-entrant sequence of first-order structural phase transitions (*Pcmb*→ *P2/c*11→ *P*12$_1$/*m*1→ *Pcmb*) that is unprecedented among brownmillerites, broadening the family of space groups available to these materials and challenging current approaches for sorting the myriad variants of brownmillerite structures. Magnetic susceptibility data indicate antiferromagnetic ordering in $Ca_2Co_2O_5$ occurs near 240 K, corroborated by neutron powder diffraction. Below 140 K, the specimen exhibits a weak ferromagnetic component directed primarily along the *b* axis that shows a pronounced thermal and magnetic history dependence.

**INTRODUCTION**

Cobalt oxides[1] attract both fundamental and technological attention due to the myriad chemically tunable physical properties they exhibit, such as thermoelectricity,[2,3] giant magnetoresistance,[4] superconductivity[5] and multiferroicity[6]. These properties can be traced to the ability of cobalt to adopt several possible oxidation states ($Co^{2+}$, $Co^{3+}$ and $Co^{4+}$), multiple types of coordination (tetrahedral, pyramidal and octahedral), and variable spin states (low spin, intermediate spin and high spin). The former two traits also result in rich crystal chemistry with various structural dimensionalities—3D, 2D, or 1D—that allow a great flexibility of the oxygen framework, making oxygen content an additional important parameter for tuning their physical properties.[1]

Perhaps unexpectedly, the pseudobinary CaO-CoO system lacks a perovskite; indeed, only two stable phases, namely $Ca_3Co_2O_6$ and $Ca_3Co_4O_9$, are identified on its structural phase diagram[7]. The former is an Ising spin chain compound exhibiting geometrical frustration,[8-13] while the latter is widely studied for its exceptional thermoelectric properties.[3,14-18] Many attempts have been made to synthesize compounds with Ca/Co=1. While the cubic perovskite $SrCoO_3$ has been synthesized by high pressure techniques,[19] the direct synthesis of the analogous $CaCoO_3$ has been elusive. For oxygen-deficient $Ca_2Co_2O_5$, four different structures are reported in literature, none of which have been definitively shown to be brownmillerite: (**i**) Rao[20] and Chowdari[21] reported that $Ca_2Co_2O_5$ can be indexed using an orthorhombic lattice ($a\sim$11.0 Å, $b\sim$10.7 Å and $c\sim$8.0 Å). Their suggested model bears resemblance to $Ca_2Mn_2O_5$, which has ordered anion vacancies in alternate (110) planes of the cubic perovskite structure. But no detailed structural information was provided. (**ii**) According to Funahashi et al.,[25] $Ca_2Co_2O_5$ has a layered structure similar to that of $Ca_3Co_4O_9$, which has 25% Ca sites vacant. It shows excellent thermoelectric properties.[22-25] In fact, this cobaltite, as well as $Ca_3Co_4O_9$ and $Ca_9Co_{12}O_{28}$, belongs to a misfit-layered cobaltite family, corresponding to the formulation $[Ca_2CoO_3][CoO_2]_{1.62}$.[1] (**iii**) Kim et al.[26] synthesized $CaCoO_{2.52}$ by the dripping pyrolysis method and indexed the powder XRD pattern with approximately twice the size of the ideal perovskite unit cell. (**iv**) Im et al.[27] obtained $CaCoO_{2.52}$ using a similar method as Kim[26] but indexed the phase as an orthorhombic system and described it as a brownmillerite[28-30] structure based solely on the lattice parameters ($a$~7.72 Å, $b$~15.80 Å, and $c$~7.86 Å), i.e., with no verifiable structural model. Substitution of trivalent metals on the Co site (i.e., $Ca_2Co_{2-x}M^{3+}_xO_5$) can indeed stabilize the brownmillerite structure; for example $Ca_2Co_{1.26}Al_{0.73}O_5$,[31] $Ca_2Co_{1.54}Ga_{0.46}O_5$,[32] and $Ca_2FeCoO_5$.[28] Recently, brownmillerite $Ca_2Co_2O_5$ has been epitaxially stabilized using $NdGaO_3$ substrate and was reported to crystallize in the space group *Ibm*2.[33] In summary, the unsubstituted bulk phase of brownmillerite $Ca_2Co_2O_5$ has remained elusive, along with its detailed crystal structure and physical properties. The successful synthesis of the brownmillerite form of $Ca_2Co_2O_5$ would allow direct comparison to the extensive understanding of the chemical, magnetic, and electronic structure of this rich family of compounds, and could also provide entre to functional platforms, such as an fuel cell electrode material (as a fast oxide-ion conductor)[34] and as dense membranes[35] for the separation of oxygen from gas mixtures.

In this contribution, we report the synthesis of single crystals of $Ca_2Co_2O_5$ using a high pressure optical-image floating zone technique. The high oxygen fugacity is critical to synthesize this compound as a pure phase and to grow single crystals. We find that as-synthesized $Ca_2Co_2O_5$ is an ordered oxygen-deficient perovskite of the brownmillerite type, and it undergoes an unprecedented re-entrant structural phase transitions (*Pcmb*→ *P*2/*c*11→ *P*12$_1$/*m*1→ *Pcmb*) with decreasing temperature. We describe its temperature-dependent structural, thermal, and magnetic properties, including antiferromagnetic ordering near 240 K. Finally, we discuss the challenge that $Ca_2Co_2O_5$ presents to the community's understanding of the structural sorting map of ordered and disordered brownmillerites and develop a semi-quantitative understanding based on the accepted model of dipole-dipole interactions between tetrahedral chains.[29,36,37]

**EXPERIMENTAL SECTION**

**Solid State Reaction**. Precursors for crystal growth were synthesized via standard solid-state reaction techniques. Stoichiometric ratios of $CaCO_3$ (Alfa Aesar, 99.999%) and $Co_3O_4$ (Alfa Aesar, 99.9985%) were thoroughly ground and then loaded into a Pt crucible. The mixture was heated in $O_2$ atmosphere from room temperature to 900 °C at a rate of 3 °C/min and allowed to dwell at the temperature for 24 h, then furnace-cooled to room temperature. The solid was then reground, and sintered twice at 1000 °C using the procedures mentioned above. Phase identification of the resultant black powder was performed by powder X-ray diffraction (PXRD, PANalytical X'Pert PRO) using Cu K$\alpha$ radiation ($\lambda$=1.54056 Å). The powder was then hydrostatically pressed into polycrystalline rods (length=100 mm, diameter=8 mm) and sintered again at 1000 °C for 24 h.

**High Pressure Crystal Growth**. Single crystals of $Ca_2Co_2O_5$ were grown in $O_2$ atmosphere at pressure of 145 bar using a vertical optical-image floating zone furnace designed for operation at elevated gas pressure (Model HKZ, SciDre GmbH, Dresden). A 5 kW Xenon arc lamp was utilized for heating the zone. During growth, a flow rate of 0.1 L/min of oxygen was maintained. Feed and seed rods were counter-rotated at 20 rpm and 15 rpm, respectively, to improve zone homogeneity. The travelling speed of the seed was 10 mm/h. After 1.5 hours of growth, the zone and boule were quenched by shutting down the lamp. PXRD measured on pulverized crystals recovered from the boule revealed a phase that could be indexed on a brownmillerite cell. A representative crystal with dimensions of 1.4×0.8×0.5 mm$^3$ is shown in **Fig. 1(*a*)**. The quality of the crystal shown in **Fig. 1(*a*)** was evaluated using X-ray diffraction, as shown in **Fig. 1(*b*)**, at beamline 11-ID-C at Advanced Photon Source at Argonne National Laboratory with an average wavelength of 0.11165 Å and beam size of 0.8 mm×0.8 mm.

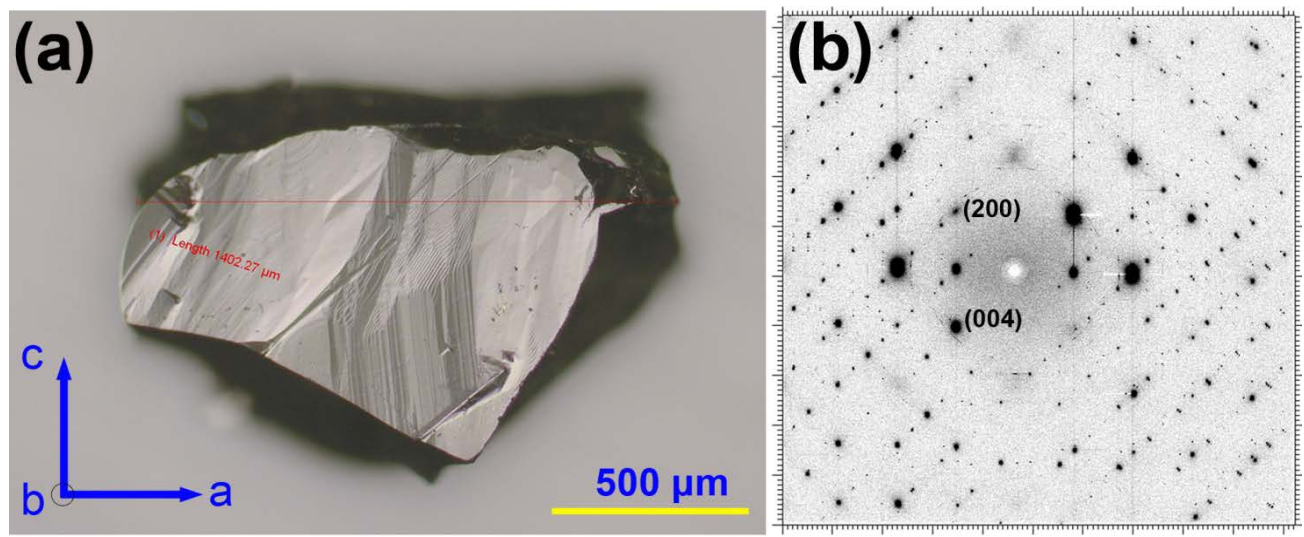

**Fig.1**. (a) As-grown $Ca_2Co_2O_5$ single crystal (2.1 mg, 1.4×0.8×0.5 mm$^3$) and (b) The X-ray diffraction pattern in the *a\*c\** plane.

**Single-Crystal Structure Determination**. A single crystal of brownmillerite $Ca_2Co_2O_5$ with dimensions of 0.05×0.05×0.06 mm$^3$ was affixed to the tip of a glass fiber and mounted on a Bruker AXS SMART three-circle diffractometer equipped with an APEX II CCD detector and Mo Kα radiation (λ=0.71073 Å). Temperature control in the 300-100 K range was provided by an Oxford Cryostream 700 Plus Cooler with uncertainties of ±0.2 K. Preliminary lattice parameters and orientation matrices were obtained from three sets of frames. Full data sets for structural analysis were collected at temperatures of 300, 240, 200 and 100 K with a detector distance of 50 mm. Data integration and cell refinement were performed by the SAINT program of the APEX2 software and multi-scan absorption corrections were applied using the SCALE program for area detector.[38] The structures were solved by direct methods and refined by full matrix least-squares methods on $F^2$. All atoms were refined with anisotropic thermal parameters, and the refinements converged for $I > 2\sigma(I)$. Calculations were performed using the SHELXTL crystallographic software package.[39] Details of crystal parameters, data collection and structure refinement at each temperature are summarized in **Table I**. Atomic positions at each temperature are listed in **Table S1**. Selected bond distances (Å) and angles (°), bond valence sums (BVS), and polyhedral distortion parameters (Δ) are presented in **Table II**. Additional information in the form of CIF has also been supplied as Supporting Information. Further details of the crystal structures may be obtained from Fachinformationszentrum Karlsruhe, 76344 Eggenstein-Leopoldshafen, Germany (fax: (+49)7247-808-666;E-mail: crysdata@fiz-karlsruhe.de, http://www.fiz-karlsruhe.de/request_for_deposited_data.html) on quoting the appropriate CSD number 428757-428760.

**Table I**. Summary of crystal data and structure refinement for $Ca_2Co_2O_5$ at various temperatures.

| Temperature | 300(2) K | 240(2) K | 200(2) K | 100(2) K |
|---|---|---|---|---|
| Empirical formula | $Ca_2Co_2O_5$ | $Ca_2Co_2O_5$ | $Ca_2Co_2O_5$ | $Ca_2Co_2O_5$ |
| Formula weight | 278.02 | 278.02 | 278.02 | 278.02 |
| Wavelength | 0.71073 Å | 0.71073 Å | 0.71073 Å | 0.71073 Å |
| Crystal system | Orthorhombic | Monoclinic | Monoclinic | Orthorhombic |
| Space group [a] | *Pcmb* (No. 57) | *P*2/*c*11(No. 13) | *P*12$_1$/*m*1(No. 11) | *Pcmb* (No. 57) |
| Unit cell dimensions | $a$ = 5.28960(10) Å  $b$ = 14.9240(2) Å  $c$ = 10.9547(2) Å | $a$ = 5.29440(10) Å  $b$ = 14.8022(3) Å  $c$ = 10.9667(2) Å  $\alpha$ = 90.0780(10)° | $a$ = 5.30280(10) Å  $b$ = 14.7838(3) Å  $c$ = 10.9810(2) Å  $\beta$ = 90.1730(10)° | $a$ = 5.30380(10) Å  $b$ = 14.7518(2) Å  $c$ = 10.9877(2) Å |
| Volume | 864.79(3) Å$^3$ | 859.45(3) Å$^3$ | 860.86(3) Å$^3$ | 859.68(3) Å$^3$ |
| Z | 8 | 8 | 8 | 8 |
| Density (calculated) | 4.271 g/cm$^3$ | 4.297 g/cm$^3$ | 4.290 g/cm$^3$ | 4.296 g/cm$^3$ |
| Absorption coefficient | 9.956 mm$^{-1}$ | 10.018 mm$^{-1}$ | 10.002 mm$^{-1}$ | 10.015 mm$^{-1}$ |
| Crystal size | 0.06×0.05×0.05 mm$^3$ | | | |
| θ range for data collection | 2.73 to 31.54° | 1.38 to 31.53° | 1.85 to 31.50° | 2.76 to 31.50° |
| Index ranges | -7<=h<=7, -21<=k<=21, -16<=l<=15 | -21<=h<=21, -7<=k<=7, -16<=l<=16 | -7<=h<=7, -21<=k<=21, -16<=l<=16 | -7<=h<=7, -21<=k<=21, -16<=l<=16 |
| Reflections collected | 11175 | 11400 | 11770 | 11131 |
| Independent reflections | 1488 [$R_{int}$ = 0.0359] | 2841 [$R_{int}$ = 0.0327] | 2939 [$R_{int}$ = 0.0310] | 1474 [$R_{int}$ = 0.0327] |
| Completeness | 99.7% | 99.3% | 99.6% | 99.8% |
| Refinement method | Full-matrix least-squares on F$^2$ | | | |
| Data / restraints / parameters | 1488 / 0 / 91 | 2841 / 0 / 168 | 2939 / 0 / 178 | 1474 / 0 / 90 |
| Goodness-of-fit | 1.000 | 1.050 | 0.984 | 1.039 |
| Final R indices [>2σ(I)] [b] | $R_{obs}$ = 0.0221  $wR_{obs}$ = 0.0620 | $R_{obs}$ = 0.0299  $wR_{obs}$ = 0.0779 | $R_{obs}$ = 0.0265  $wR_{obs}$ = 0.0655 | $R_{obs}$ = 0.0235  $wR_{obs}$ = 0.0605 |
| R indices [all data] | $R_{all}$ = 0.0374  $wR_{all}$ = 0.0703 | $R_{all}$ = 0.0521  $wR_{all}$ = 0.0871 | $R_{all}$ = 0.0528  $wR_{all}$ = 0.0760 | $R_{all}$ = 0.0382  $wR_{all}$ = 0.0682 |
| Largest diff. peak and hole | 0.735 and -0.936 e·Å$^{-3}$ | 0.969 and -0.954 e·Å$^{-3}$ | 0.803 and -0.958 e·Å$^{-3}$ | 0.638 and -1.471 e·Å$^{-3}$ |

[a] The non-standard settings (*Pcmb* and P2/*c*11) are adopted so that a uniform setting b>c>a is maintained following the convention of Ref. 28.

[b] $R = \Sigma||F_o|-|F_c|| / \Sigma|F_o|$, $wR = \{\Sigma[w(|F_o|^2 - |F_c|^2)^2] / \Sigma[w(|F_o|^4)]\}^{1/2}$ and calc w=1/[$\sigma^2(F_o^2)$+(AP)$^2$+BP] where P=($F_o^2$+2$F_c^2$)/3, A/B are 0.0379/0.0820 (300 K), 0.0367/0.0000 (240 K), 0.0362/1.7107 (200 K), and 0.0403/0.0000 (100 K).

**Table II**. Selected bond distances (Å), angles ($^0$), bond valence sum (BVS), and distortion parameter ($\Delta$) for $Ca_2Co_2O_5$ at various temperatures based on single crystal X-ray diffraction.

|  | 300 K |  | 240 K |  | 200 K |  | 100 K |  |
|---|---|---|---|---|---|---|---|---|
| **Co tetrahedra** | Co(1)-O(4) | 1.799(5) | Co(1)-O(3) | 1.782(4) | Co(1)-O(7) | 1.806(3) | Co(1)-O(4) | 1.800(3) |
|  | Co(1)-O(4) | 1.799(5) | Co(1)-O(4) | 1.818(5) | Co(1)-O(7) | 1.806(3) | Co(1)-O(4) | 1.801(3) |
|  | Co(1)-O(2) | 1.930(3) | Co(1)-O(1) | 1.921(3) | Co(1)-O(3) | 1.922(3) | Co(1)-O(2) | 1.924(2) |
|  | Co(1)-O(1) | 1.934(2) | Co(1)-O(2) | 1.928(2) | Co(1)-O(1) | 1.918(3) | Co(1)-O(1) | 1.925(2) |
| BVS$^a$ |  | 2.60 |  | 2.62 |  | 2.61 |  | 2.61 |
| $\Delta$(Co-O)$^b \times 10^4$ |  | 12.7 |  | 11.7 |  | 9.4 |  | 11.1 |
|  |  |  |  |  | Co(2)-O(8) ×2 | 1.806(3) |  |  |
|  |  |  |  |  | Co(2)-O(4) | 1.926(3) |  |  |
|  |  |  |  |  | Co(2)-O(2) | 1.924(3) |  |  |
| BVS |  |  |  |  |  | 2.59 |  |  |
| $\Delta$(Co-O)$\times 10^4$ |  |  |  |  |  | 10.2 |  |  |
|  | Co(2)-O(3) | 1.801(4) | Co(2)-O(5) | 1.792(4) | Co(3)-O(6) | 1.792(3) | Co(2)-O(3) | 1.802(3) |
|  | Co(2)-O(3) | 1.801(4) | Co(2)-O(10) | 1.815(4) | Co(3)-O(6) | 1.792(3) | Co(2)-O(3) | 1.802(3) |
|  | Co(2)-O(1) | 1.923(3) | Co(2)-O(2) | 1.916(3) | Co(3)-O(2) | 1.926(3) | Co(2)-O(1) | 1.916(3) |
|  | Co(2)-O(2) | 1.936(2) | Co(2)-O(1) | 1.930(3) | Co(3)-O(4) | 1.933(3) | Co(2)-O(2) | 1.926(2) |
| BVS |  | 2.60 |  | 2.61 |  | 2.64 |  | 2.62 |
| $\Delta$(Co-O) $\times 10^4$ |  | 11.9 |  | 10.5 |  | 13.7 |  | 10.3 |
|  |  |  |  |  | Co(4)-O(5) ×2 | 1.799(3) |  |  |
|  |  |  |  |  | Co(4)-O(1) | 1.920(3) |  |  |
|  |  |  |  |  | Co(4)-O(3) | 1.930(3) |  |  |
| BVS |  |  |  |  |  | 2.62 |  |  |
| $\Delta$(Co-O) $\times 10^4$ |  |  |  |  |  | 11.5×10$^{-4}$ |  |  |
| Tetrahedral twist angle $^c$ | O(1)-O(2)-O(1) | 121.5(1) | O(2)-O(1)-O(2) | 122.2(1) | O(1)-O(3)-O(1) | 122.4(1) | O(1)-O(2)-O(1) | 122. 8(1) |
|  |  |  |  |  | O(2)-O(4)-O(2) | 122.1(1) |  |  |
| Layer separation |  | 7.46 |  | 7.40 |  | 7.39 |  | 7.38 |
| **Co octahedra** | Co(3)-O(5) ×2 | 1.894(4) | Co(3)-O(8) ×2 | 1.887(3) | Co(5)-O(9) ×2 | 1.879(3) | Co(3)-O(5) ×2 | 1.872(2) |
|  | Co(3)-O(6) ×2 | 1.932(4) | Co(3)-O(9) ×2 | 1.944(3) | Co(5)-O(11) ×2 | 1.957(3) | Co(3)-O(6) ×2 | 1.961(2) |
|  | Co(3)-O(3) ×2 | 2.114(4) | Co(3)-O(10) ×2 | 2.061(4) | Co(5)-O(5) ×2 | 2.090(3) | Co(3)-O(3) ×2 | 2.085(3) |
| BVS |  | 2.91 |  | 2.99 |  | 2.93 |  | 2.95 |
| $\Delta$(Co-O) $\times 10^4$ |  | 23.5 |  | 13.6 |  | 19.4 |  | 19.6 |
|  |  |  | Co(4)-O(6) ×2 | 1.876(3) | Co(6)-O(10) ×2 | 1.871(3) |  |  |
|  |  |  | Co(4)-O(7) ×2 | 1.962(3) | Co(6)-O(12) ×2 | 1.983(3) |  |  |
|  |  |  | Co(4)-O(5) ×2 | 2.119(4) | Co(6)-O(6) ×2 | 2.103(3) |  |  |
| BVS |  |  |  | 2.87 |  | 2.86 |  |  |
| $\Delta$(Co-O) $\times 10^4$ |  |  |  | 25.7$^4$ |  | 22.8 |  |  |
|  | Co(4)-O(5) | 1.907(4) | Co(5)-O(8) | 1.911(3) | Co(7)-O(9) | 1.901(3) | Co(4)-O(5) | 1.919(2) |
|  | Co(4)-O(5) | 1.907(4) | Co(5)-O(8) | 1.911(3) | Co(7)-O(10) | 1.925(3) | Co(4)-O(5) | 1.919(2) |
|  | Co(4)-O(6) | 1.916(4) | Co(5)-O(9) | 1.917(3) | Co(7)-O(11) | 1.930(3) | Co(4)-O(6) | 1.922(2) |
|  | Co(4)-O(6) | 1.916(4) | Co(5)-O(9) | 1.917(3) | Co(7)-O(12) | 1.897(3) | Co(4)-O(6) | 1.922(2) |
|  | Co(4)-O(4) | 2.110(4) | Co(5)-O(3) | 2.124(5) | Co(7)-O(7) | 2.072(3) | Co(4)-O(4) | 2.072(3) |
|  | Co(4)-O(4) | 2.110(4) | Co(5)-O(3) | 2.124(5) | Co(7)-O(8) | 2.075(3) | Co(4)-O(4) | 2.072(3) |
| BVS |  | 2.92 |  | 2.88 |  | 2.98 |  | 2.94 |

| | | | | |
|---|---|---|---|---|
| Δ(Co-O) ×10⁴ | 22.4 | 24.9 | 15.1 | 13.1 |
| | Co(6)-O(6) ×2 | 1.909(3) | | |
| | Co(6)-O(7) ×2 | 1.911(3) | | |
| | Co(6)-O(4) ×2 | 2.031(5) | | |
| BVS | | 3.08 | | |
| Δ(Co-O) ×10⁴ | | 8.6 | | |

[a] $BVS = \sum_{i=1}^{N} v_i$, $v_i = exp[(R_0 - d_n)/B]$, N is the coordination number, $B=0.37$, $R_0(Co^{3+})=1.70$;[40,41]
[b] $\Delta(Co-O) = (1/N) \sum_{n=1}^{N} \{(d_n - d_{av})/d_{av}\}^2$, $d_{av} = (1/N) \sum_{i=1}^{N} d_i$ is the average Co-O distance;[42]
[c] Definitions of tetrahedral twist and layer separation, see text and *Ref. 29*.

**High Resolution Variable-Temperature PXRD (HRXRD)**. HRXRD data were collected at beamline 11-BM of the Advanced Photon Source (APS) at Argonne National Laboratory in the range $0.5° \leq 2\theta \leq 34°$ with a step size of $0.001°$ and wavelength $\lambda=0.41367$ Å. A powder sample was prepared by loading pulverized single crystals into a Φ0.3 mm quartz capillary that was spun continuously at 560 rpm during data collection. Diffraction patterns were recorded on cooling at 300, 280, 260, 250, 246-220 (2 K interval), 215, 210-100 K (10 K interval). An Oxford Cryostream 700 plus $N_2$ gas blower was used to control temperature, and the cooling rate was set to 2 K/min for 300- 246 K, 1 K/min for 246-220 K, and 2 K/min for 215-100 K. Temperature was stabilized for five min at each set point prior to data collection. The obtained HRXRD data were analyzed using the GSAS[43] software with the graphical interface EXPGUI[44] using the respective single crystal structural models as starting points. Refined parameters include background, intensity scale factor, $2\theta$ zero offset, lattice parameters, atomic positions (except oxygen), isotropic atomic displacement parameters ($U_{iso}$, grouped by atomic species), and profile shape parameters. Shifted Chebyshev and pseudo-Voigt functions with anisotropic microstrain broadening terms (function #4)[45] were used for the background and peak profiles, respectively. For two-phase refinements, the profile parameters were constrained to be the same in each phase. The Gaussian terms (GU, GV and GW) were also refined to account for the observed peak shape.

**Magnetic Susceptibility**. Magnetic susceptibility measurements were performed on single crystals (aligned at beamline 11-ID-C at the APS) using a Quantum Design SQUID magnetometer. The crystal shown in **Fig. 1** was attached to a quartz rod using a minute amount of GE-varnish. Zero-field cooled (ZFC) and field cooled (FC) data along crystallographic *a*-, *b*-, and *c*- axis were collected between 1.8 and 400 K under an external field of 0.01 T. ZFC and FC data along the *b* axis were also recorded under external fields of 0.02, 0.05, 0.1, and 7 T. In the ZFC protocol the sample was cooled in zero field to 10 K at a rate of 10 K/min and then to 1.8 K at a rate of 2 K/min, and dc magnetization recorded while warming (2 K/min from 1.8 to 90 K, 0.5 K/min from 90 to 260 K, and 2 K/min from 260 to 400 K). In the FC protocol the magnetization was recorded while cooling (2 K/min from 400 to 260 K, 0.5 K/min from 260 to 90 K, and 2 K/min from 90 to 1.8 K) in a fixed field. Isothermal field-dependent magnetization at 10 K along *a*-, *b*-, and *c*-axis was measured in a field range of ±7 T following field cooling (10 K/min) under 7 T. Isothermal field-dependent magnetization at 10 K along *b*-axis was also measured using the following three protocols: (i) slow cooling (2 K/min from above room temperature to 260 K, 0.5 K/min from 260 to 90 K, then 2 K/min from 90 to 10 K) under 7 T; (ii) fast cooling (10 K/min) under 0 T; (iii) slow cooling (2 K/min from above room temperature to 260 K, 0.5 K/min from 260 to 90 K, then 2 K/min from 90 to 10 K) under 0 T.

**Differential Scanning Calorimetry (DSC).** DSC for $Ca_2Co_2O_5$ between room temperature and 1273 K was carried out on a NETZSCH DSC 404 analyzer with ~100 mg fine powders placed in an $Al_2O_3$ crucible and heated at a rate of 10 K/min in air. After the measurements, the residue was analyzed by powder X-ray diffraction. Low temperature (123- 288 K) DSC was performed on a Mettler Toledo DSC 823$^e$. 15.335 mg pulverized crystals were loaded in an Al pan and cooled/heated at a rate of 5 K/min.

**Thermogravimetric analysis (TGA).** Oxygen content of $Ca_2Co_2O_5$ was determined by reduction in a 4% $H_2$/Ar mixture on a thermogravimetric analysis balance (Mettler-Toledo Model TGA/DSC 1). A powder sample (~100 mg) was placed into a 150 microliter $Al_2O_3$ crucible and heated at a rate of 10 °C/min from room temperature to 900 °C, and then held isothermally for five hours.

## RESULTS AND DISCUSSION

**Synthesis and High Pressure Crystal Growth**. The PXRD pattern of the precursor powders (see **Fig. S1**) shows a mixture of $Ca_3Co_2O_6$ and $Ca_3Co_4O_9$ consistent with the published CaO-CoO phase diagram,[7] which shows no brownmillerite $Ca_2Co_2O_5$ at atmospheric pressure. However, a new phase was obtained when a mixed-phase precursor rod was melted and quenched under conditions of high $pO_2$ (145 bar). Its PXRD pattern (see **Fig. S2**), matches well with that of $Ca_2Co_{1.54}Ga_{0.46}O_5$,[32] indicating formation of a brownmillerite phase.

To obtain single crystals of the brownmillerite phase, growth conditions such as oxygen pressure, rotation speed, travelling speed, power, and post-growth cooling rate, were explored. We found that $pO_2$ indeed is a key factor for $Ca_2Co_2O_5$ formation, with $pO_2$ ≥50 bar required to obtain high-purity $Ca_2Co_2O_5$ crystals (see **Fig. S3**). For growth at $pO_2$ < 50 bar, a mixture of CoO, CaO and $Ca_2Co_2O_5$ was obtained. Despite multiple attempts under varying conditions, zone instability (perhaps resulting from strong convection in the high pressure ambient) precluded long growth cycles, and we have not yet succeeded in growing large single crystals often obtained in floating zone furnaces. Nonetheless, sizable single crystals can be cleaved from the as-grown boule (see **Fig. 1(a)**) following rapid cooling of the specimen in the zone furnace. This rapid cooling is necessary because $Ca_2Co_2O_5$ decomposes to $Ca_3Co_4O_9$ and other unidentified compounds during the post-growth cooling process even at $pO_2$ as high as 145 bar. The diffraction pattern of the cleaved crystal, shown in **Fig. 1(b),** demonstrates that it is a high quality crystal specimen.

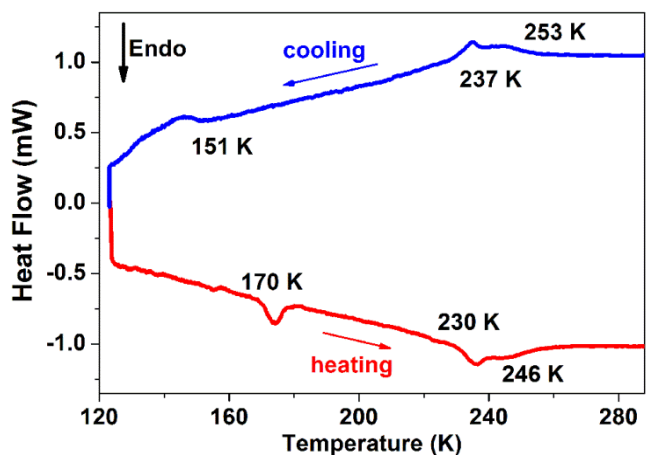

**Fig. 2.** DSC data between 123 K and 288 K for $Ca_2Co_2O_5$. The extrapolated onset temperatures are indicated. The arrows indicate cooling and warming process.

**Thermal Analysis**. Four batches of pulverized crystals were used to establish the O content in $Ca_2Co_2O_5$ via TGA, which was determined to be 5.00±0.03. To evaluate the thermal stability of $Ca_2Co_2O_5$, DSC was performed on a pulverized single crystal sample between room temperature and 1273 K. The DSC curve reveals that $Ca_2Co_2O_5$ retains the brownmillerite structure in air to ~570 °C, reflecting a lower stability than that of $Sr_2Co_2O_5$ (653 °C)[46]. The PXRD pattern of the residue in the $Al_2O_3$ crucible after DSC is a combination of $Ca_3Co_2O_6$ and $Ca_3Co_4O_9$, consonant with the ambient pressure CaO-CoO phase diagram.[7] Low temperature DSC curves are shown in **Fig. 2**. Three exothermic peaks, located at 253, 237 and 151 K, are clearly seen on the cooling curve, and three corresponding endothermic peaks are observed at 246, 230 and 170 K on the warming curve. These results demonstrate that $Ca_2Co_2O_5$ experiences three first-order phase transitions between 123 K and room temperature.

**Crystal Structure and Re-entrant Phase Transition**. With decreasing temperature, the $Ca_2Co_2O_5$ single crystal undergoes a series of structural phase transitions that are heretofore unreported among brownmillerites. Specifically, $Ca_2Co_2O_5$ transforms from a room-temperature orthorhombic form (space group *Pcmb,* alternate setting of *Pbcm*, No. 57) to an intermediate monoclinic phase (space group $P2/c11$) at 253 K, to a second intermediate monoclinic phase (space group $P12_1/m1$) at 237 K, and then finally re-enters the orthorhombic *Pcmb* phase at 151 K.

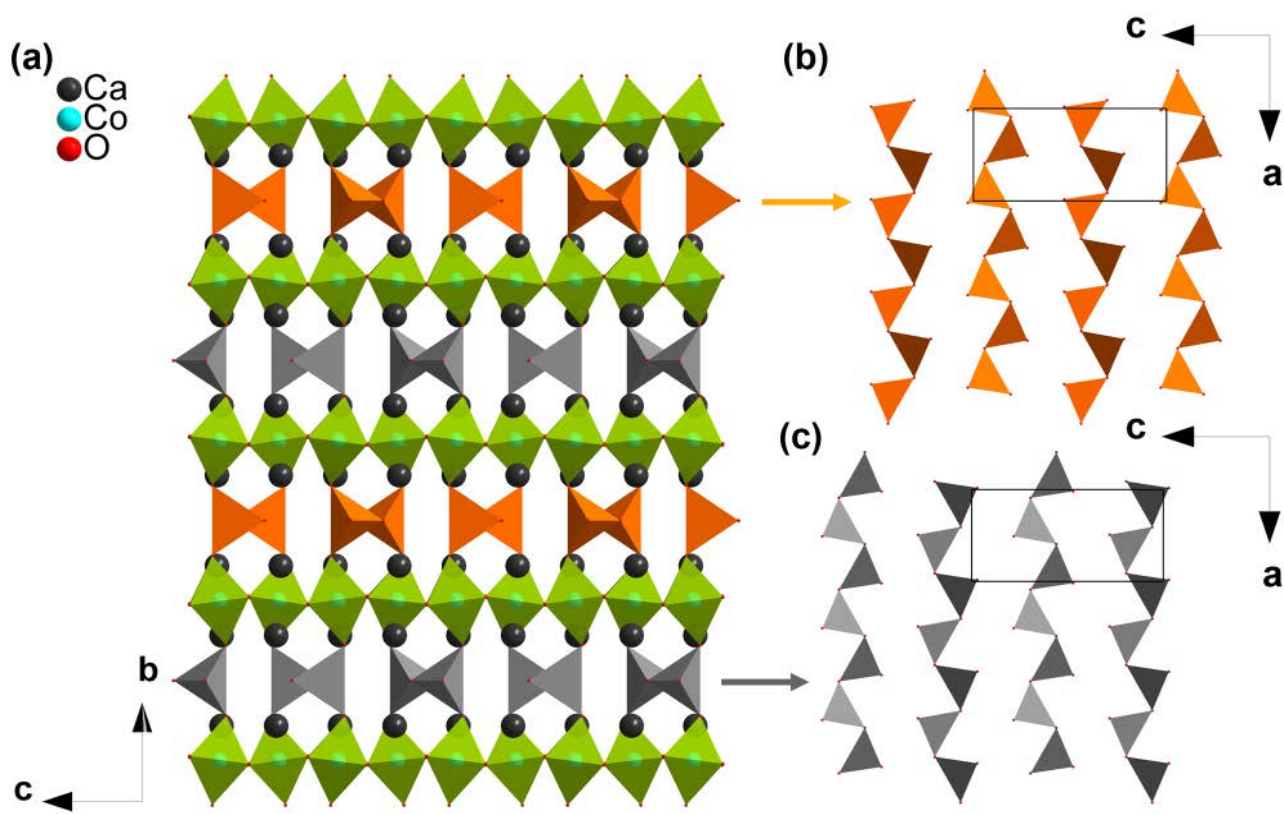

**Fig.3**. $Ca_2Co_2O_5$ structure (a) polyhedral representation in the *bc*-plane; (b) Tetrahedral chain orientation at y=0.75; (c) Tetrahedral chain orientation at y=0.25.

**Fig. 3** shows the structure of $Ca_2Co_2O_5$ at 300 K. It belongs to the orthorhombic space group *Pcmb* with unit cell parameters *a*=5.28960(10) Å, *b*=14.9240(2) Å, *c*=10.9547(2) Å, and *Z*=8. It is worth noting that oxides crystallizing in this space group with one short axis doubled (*c* axis in this setting) are exceedingly rare among brownmillerites, having been reported to our knowledge only for

$Ca_2FeCoO_5$[28] and $La_{2-x}A_xMn_2O_5$ (A=Ca or Sr, $x$=0.7 and 0.8)[29]. In $Ca_2Co_2O_5$, the Co cations are completely ordered with full occupancy at all sites. In addition, both intra- and interlayer ordering of tetrahedral chain orientations are observed, i.e., the left- and right-handed orientations alternate within each tetrahedral layer as well as between the closest neighboring tetrahedral layers. This ordering can be seen in **Figs. 3(b&c)**, where the tetrahedral chains propagating along the *a* axis have opposite orientations relative to the neighboring chains within the same layer (grey or orange) and in the adjacent layer (grey and orange). There are two Ca atoms, four Co atoms, and six O atoms in the asymmetric unit of $Ca_2Co_2O_5$. The four symmetry-independent Co atoms fall into two sets: two in octahedral coordination with Co-O bond distances ranging from 1.894(4) to 2.114(4) Å and two in tetrahedral environments with Co-O bond distance of 1.799(5)- 1.936(2) Å (See **Table II**). Ca atoms are found between the layers to maintain charge balance.

The single crystal X-ray data collected at 240 K can be indexed in the monoclinic crystal system with unit cell dimensions of $a$=5.29440(10) Å, $b$=14.8022(3) Å, $c$=10.9667(2) Å, and $\alpha$ = 90.0780(10)°. The program XPREP[38] suggested the space group $P2_1/c11$ (No. 14), but the refinement converged to an unsatisfactory $R_{obs}$ = 0.0896 with disordered cobalt and oxygen positions. The common monoclinic structure found in perovskites, $C2/c11$ (No. 15), is ruled out due to many violators of the *C* centering extinction condition. Hayward *et al.* have argued that charge ordering in the octahedral layers is responsible for stabilizing the $C2/c$ phase.[29] With no suggestion of mixed valence in $Ca_2Co_2O_5$ it is unsurprising that $C2/c11$ is not found. Following a descent of symmetry from *Pcmb*, we attempted the subgroup $P2/c11$ (No. 13) as a structural model. The atomic positions, derived from the high-temperature *Pcmb* model by disregarding the symmetry along *b* and *c* axes, were used as a starting point to solve the structure. An ordered structural model with four Ca atoms, six Co atoms and ten O atoms in the asymmetric unit was obtained, and the satisfactory refinement figures of merit ($R_{obs}$=0.0299 and $wR_{obs}$=0.0779) and small residual (0.969 and -0.954 e·Å$^{-3}$) support the choice of this model. The underlying structural framework remains the same as that of 300 K, with only subtle atomic rearrangements. The $P2/c11$ model was also used to successfully refine the high resolution PXRD data (vide infra) at 240 K, which contains a small amount of the *Pcmb* parent as a second phase.

The crystal structure of $Ca_2Co_2O_5$ at 200 K could be solved easily in the space group $P12_1/m1$ (a subgroup of index two of *Pcmb*) by direct methods, yielding the positions of four Ca atoms, seven Co atoms, and twelve O atoms. No suggestion of disorder of either Co or O atoms was found, and all of the positions are fully occupied. The structure also belongs to the brownmillerite type with interlayer, intralayer and B-site cation ordering the same as at 240 K and 300 K, shown in **Fig. 3**. The transformation from $P2/c11$ to $P12_1/m1$ at ~237 K involves the unique axis switching from the ~5 Å axis to the ~15 Å axis. To our knowledge, $Ca_2Co_2O_5$ is the first example of a brownmillerite oxide crystallizing in either of these two monoclinic space groups, $P2/c11$[47] or $P12_1/m1$, expanding the possible family of symmetries adopted by this large class of oxygen-deficient perovskites.

Finally, at 100 K the structure of $Ca_2Co_2O_5$ was solved again in *Pcmb*. In this re-entrant phase, the positions of two Ca atoms, four Co atoms, and six O atoms were obtained. Final refinement was converged to $R_{obs}$ = 0.0235 and $wR_{obs}$ = 0.0605. Despite the counterintuitive increase in symmetry on cooling, it is straightforward to visualize the re-entrant phase transformation from ordered monoclinic $P12_1/m1$ to ordered orthorhombic *Pcmb* without any substantial rearrangement. Indeed, the atomic coordinates of Ca, Co, and O atoms differ only slightly from those at 300 K (see **Table S1**)

The origin of the re-entrant structural transition in $Ca_2Co_2O_5$ is unclear. Since the structures of $Pcmb$, $P2/c11$ and $P12_1/m1$ phases are so close to each other, the difference in lattice energy must be subtle, and we might expect that entropic terms in the free energy become important. Our preliminary neutron powder diffraction data[48] collected at 200 and 130 K differ considerably, signaling changes of magnetic structure. Since this magnetic transition falls into the same temperature range as the re-entrant structural transition, it is possible that the spin re-orientation is related to this re-entrant structural transition. Further work – both experimental and theoretical – will be required to explain this unusual observation.

Re-entrant structural transitions such as that we find in $Ca_2Co_2O_5$ as a function of temperature in the absence of external stimulus (e.g., pressure, composition, magnetic field, etc.) are commonly found in liquid crystals;[49] however, they are quite rare in crystalline materials. Some well documented examples include: Rochelle salt $NaKC_4H_4O_6 \cdot 4H_2O$ ($P2_12_12 \rightarrow P2_111 \rightarrow P2_12_12$)[50], malonitrile $CH_2(CN)_2$ ($P2_1/n \rightarrow P\text{-}1 \rightarrow P2_1/n$)[51], $Li_2TiO_3$ ($Fm\text{-}3m \rightarrow C2/c \rightarrow Fm\text{-}3m$)[52], $Ag_7P_3S_{11}$ ($C2/c \rightarrow \beta\text{-}Ag_7P_3S_{11}$, unknown structure$\rightarrow C2/c$)[53], and $BaFe_2(PO_4)_2$ ($R\text{-}3 \rightarrow P\text{-}1 \rightarrow R\text{-}3$)[54]. Recently, $Rb_2SnCu_3F_{12}$ has been reported to be an additional possible case since the re-entrant character ($R\text{-}3 \rightarrow P\text{-}1 \rightarrow R\text{-}3$) was only observed in the powder form but not seen in the single crystal.[55] To explain the re-entrant structural transitions in $CH_2(CN)_2$, Rae and Dove suggested a model based on the interaction between the order parameter associated with the transitions and strains in the unit cell caused by thermal expansion.[56] Whether such an explanation can be generalized to $Ca_2Co_2O_5$ and other reentrant transitions remains an open question.

**Structural Evolution of $Ca_2Co_2O_5$ polymorphs**. The unit cell parameters, as seen from **Table I**, evolve reciprocally with increasing temperature, i.e., the $b$ axis expands while both $a$ and $c$ axes contract. The magnitudes of thermal expansion are 1.17%, -0.27% and -0.30%, respectively, from 100 K to 300 K. Notably, a minimum in the cell volume is observed at 240 K, indicating occurrence of narrow window of negative volume thermal expansion from 200 to 240 K.

Co-O bond lengths, as shown in **Table II**, fall in the range of 1.782(4)- 2.124(5) Å, comparable to other cobaltites.[1] The shortest and longest Co-O bond distances are both found in the $P2/c11$ phase, indicating this phase has the largest distortion of the polyhedral building blocks. A measure of the polyhedral distortion ($\Delta$) is calculated using $\Delta(Co-O) = (1/N)\sum_{n=1}^{N}\{(d_n - d_{av})/d_{av}\}^2$, where $d_{av} = (1/N)\sum_{i=1}^{N} d_i$ is the average Co-O distance.[42] The distortion values are presented in **Table II**. By comparison, the distortion of cobalt polyhedra in $Ca_2FeCoO_5$[28], $Ca_2Co_{1.54}Ga_{0.46}O_5$[32] and $Sr_2Co_2O_5$[57] were calculated and listed in **Table S2**. Results show that the polyhedral distortion in $Ca_2Co_2O_5$ at 300 K is comparable to that of $Ca_2FeCoO_5$. Compared with $Ca_2Co_{1.54}Ga_{0.46}O_5$ and $Sr_2Co_2O_5$, $Ca_2Co_2O_5$ has a larger distortion in tetrahedral but a smaller distortion in octahedral sites.

Bond-valence sum (BVS) calculations[40,41] indicate that the oxidation states of Ca and octahedrally coordinated Co are 1.99-2.04 and 2.86-3.08, respectively, both of which are acceptably close to the corresponding expected values based on ionic charge. In contrast, the BVS of Co ions in tetrahedral sites (2.59-2.64) deviate substantially from the nominal expectation, reflecting a considerable degree of underbonding in the tetrahedral layer. It should be noted that such cobalt underbonding in tetrahedral site is also observed in other brownmillerite oxides, such as $Ca_2FeCoO_5$[28], $Ca_2Co_{1.54}Ga_{0.46}O_5$[32] and $Sr_2Co_2O_5$[57].

To place our results into a broader context, we note that Parsons et al.[29] have proposed a structure-preference map for brownmillerites, with tetrahedral interlayer separation and the degree of twisting of the tetrahedral chains as sorting parameters. For $Ca_2Co_2O_5$, the tetrahedral chain distortion angles and interlayer separations are $121.5(1)^0$ and 7.46 Å for 300 K, $122.2(1)^0$ and 7.40 Å for 240 K, $122.4(1)^0/122.1(1)^0$ and 7.39 Å for 200 K, and $122.8(1)^0$ and 7.38 Å for 100 K (see **Table II**). Thus, $Ca_2Co_2O_5$ lies squarely in the *Pnma* region of the map, joining $Ca_2FeCoO_5$[28] as a violator of this sorting ansatz and indicating that the twist angle alone may be insufficient to parametrize the dipole moment of the tetrahedral chain. To explore why $Ca_2Co_2O_5$ falls outside the prediction of the Parsons approach, we note that compounds[29] in Parsons' data that contain A-site cations with large ion radius (La, Sr and Ba), have interchain separation larger than 7.8 Å, while those containing Ca on the A-site have an interchain separation smaller than 7.5 Å. One exception is $Ca_2Fe_{1.04}Mn_{0.96}O_5$[58]-which has an interchain separation of 7.66 Å, but it contains the Jahn-Teller (JT) active cation $Mn^{3+}$, which dominates the interlayer separation. Without evidence of JT-active ions in $Ca_2Co_2O_5$, it is easy to understand that $Ca_2Co_2O_5$ has a short interchain separation, < 7.5 Å. So the question remains: why does $Ca_2Co_2O_5$ crystallize in the *Pcmb* space group rather than *Pnma*, *I2mb* or *Imma*?

According to Abakumov,[36] Hadermann,[37] and Parsons et al.,[29] the adoption of various structural types for brownmillerite oxides principally reflects a drive by the structure to minimize the net dipole moment generated by the twisting of the tetrahedral within a chain. Three dipole cancellation schemes have been proposed: (i) When the dipole moment of each chain is small, the noncentrosymmetric space group *I2mb* is adopted. Small polar domains are suggested to form and align in an antiparallel manner to eliminate a macroscopic polarization. (ii) As the net dipole moment of each chain increases, the micro-domain cancellation scheme is replaced by an interlayer cancellation mechanism associated with the *Pnma* structure. In this case, any net polarization in one disordered tetrahedral layer is compensated on average by that of another layer. (iii) When the distance between tetrahedral layers becomes too large, interlayer cancellation becomes less favorable than an intralayer mode. In this scenario, *Pcmb* and *C2/c* structure types are observed, in which alternating chains within any given tetrahedral layer have oppositely directed polarity, leading to zero net polarization.

We attempted to extend the model of Parsons et al.[29] by calculating the dipole moments of individual tetrahedra and the chains from which they are built. Details of the calculation can be found in the Supplemental Material, and **Table S3** presents our results. Our analysis demonstrates that when the interlayer separation is less than 7.5 Å, the *Pnma* structural variant is favorable if the net chain dipole moment is less than 2.0 Debye, while *Pcmb* is preferred if the net dipole moment exceeds 2.0 Debye. This model successfully sorts the structures of six out of seven phases in **Table S3**, including violators of the Parson's scheme $Ca_2FeCoO_5$ and $Ca_2Co_2O_5$. The lone exception is $Ca_2Co_{1.54}Ga_{0.46}O_5$, which has been determined to crystallize in *Pnma* by neutron powder differaction.[32] This level of agreement argues for adopting a picture of local chemistry driving tetrahedral distortions, which in turn determine the long-range symmetry of the brownmillerites through the dipole mechanism. Thus, the local moment construct, coupled with the twist-angle ansatz proposed by Parsons et al., offers a more complete picture of the sorting mechanism that can be ultimately be traced to chemical consitutents. Clearly, additional data in this region of the sorting diagram would be desirable to provide a more rigorous test of our model.

**High Resolution Variable-Temperature PXRD.** To further investigate the re-entrant structural phase transition, *in-situ* high-resolution synchrotron X-ray powder diffraction data were collected at the beamline 11-BM of the APS. These measurements amplify the results of the single crystal studies and highlight the first-order nature of the complex phase evolution with temperature.

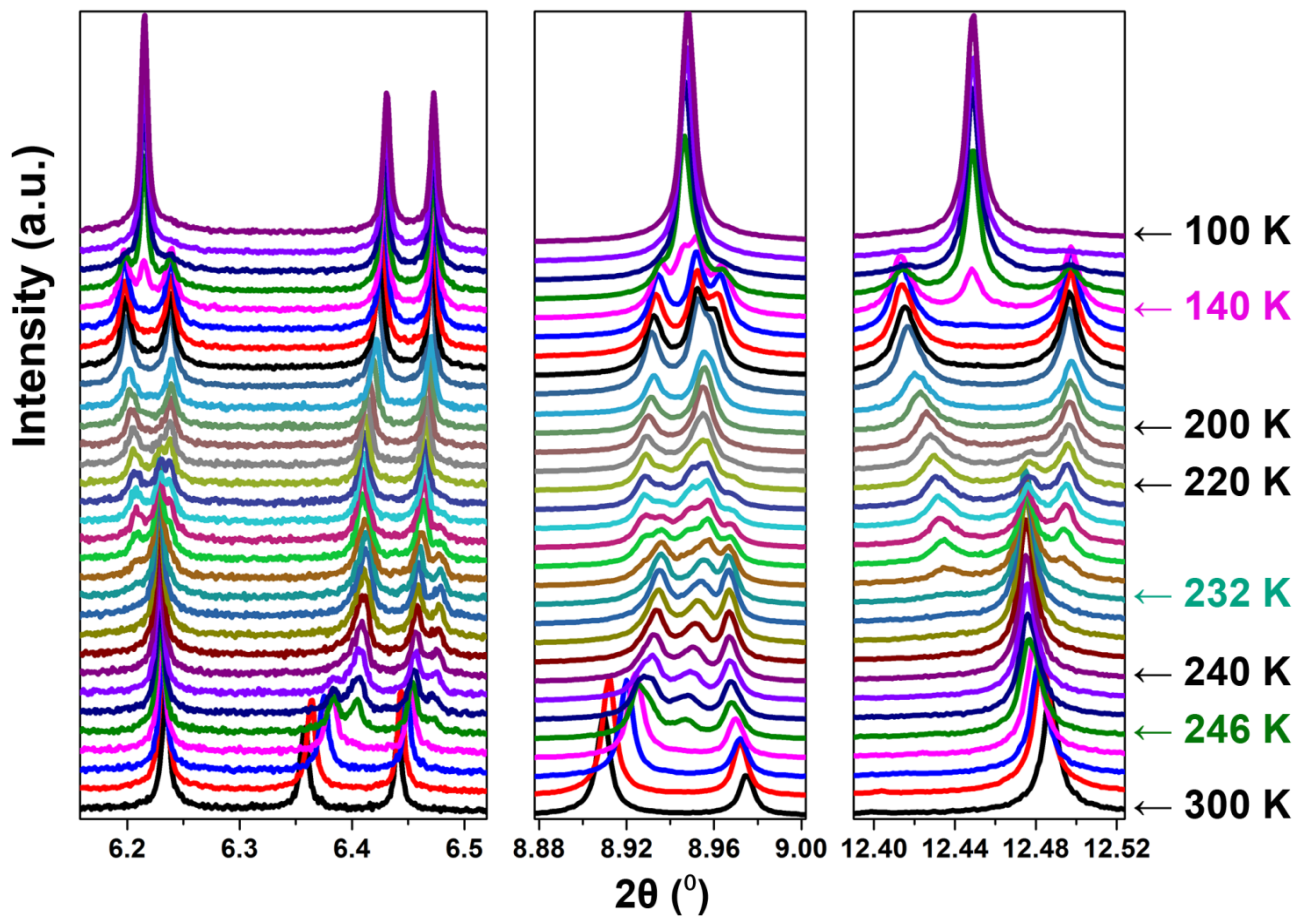

**Fig.4**. Portion of synchrotron X-ray powder diffraction patterns for $Ca_2Co_2O_5$ measured on cooling. Note: temperatures between 300 K and 246 K are 280, 260 and 250 K; between 246 and 220 K the temperature interval is 2 K; the two temperatures between 220 and 200 are 215 K and 210 K; below 210 K, the temperature interval is 10 K.

**Fig. 4**. shows selected reflections at around $6.2^0$, $6.4^0$, $8.94^0$ and $12.48^0$ at various temperatures. The evolution of these diffraction peaks as well as coexistence regions clearly establishes the three first-order phase transitions on cooling. With decreasing temperature from 300 K, peaks from a second phase are clearly seen at 246 K, 232 K and 140 K. This indicates that the three phase transitions occur above these temperatures, which is consistent with the DSC data.

We refined the patterns collected at 300 K, 240 K, 200 K and 100 K. The histograms and the results of the Rietveld refinements are shown in **Fig. 5.** The refined unit cell parameters, phase fraction as well as figures of merit ($R_{wp}$, $R_p$, $\chi^2$) are listed in **Table S4**. The 300 K and 200 K patterns were fit as single phase orthorhombic *Pcmb* and monoclinic $P12_1/m1$, respectively. The refinements converged to $R_{wp}$ =13.46% for 300 K data and $R_{wp}$ =12.22% for 200 K data.

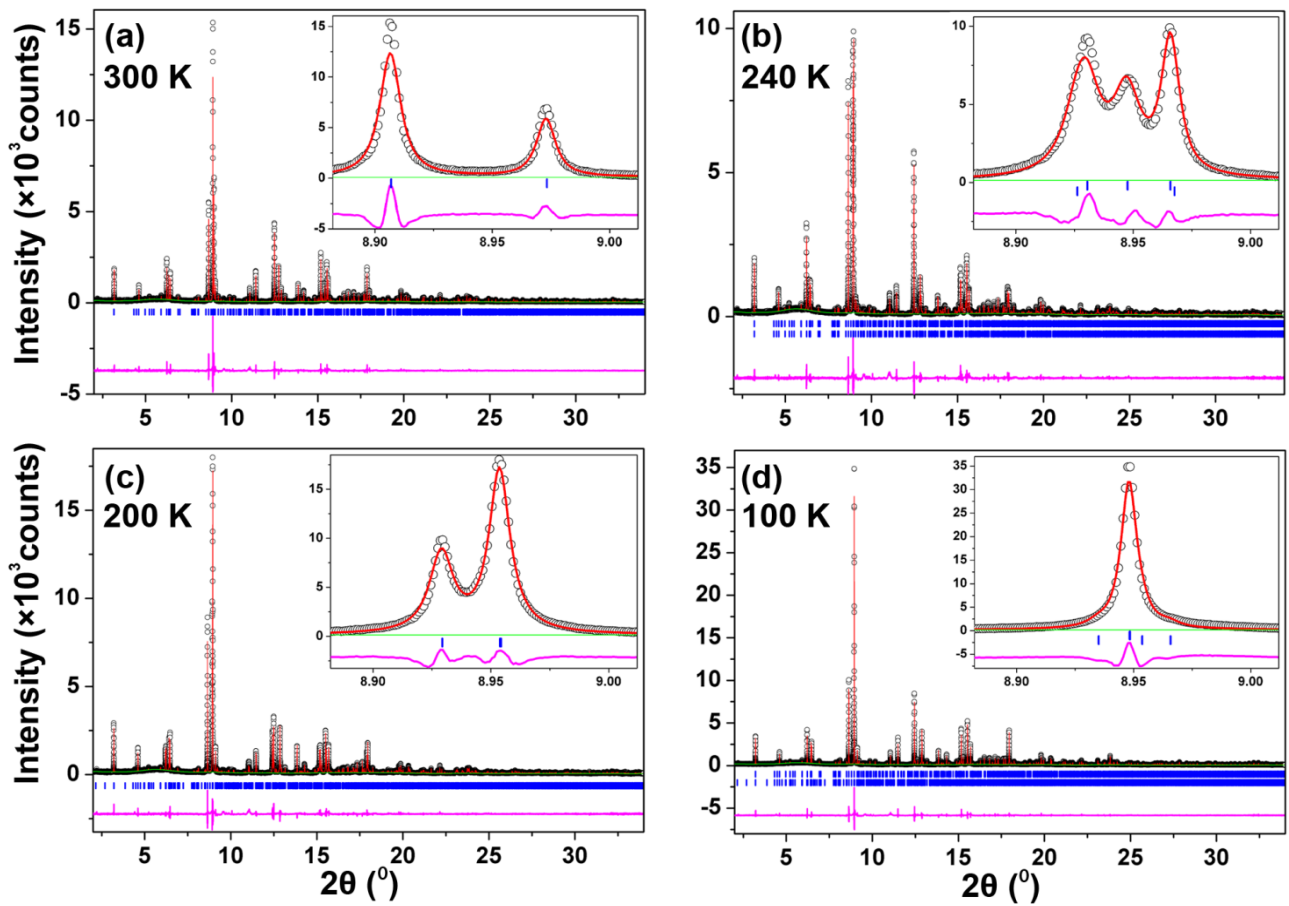

**Fig.5**. High resolution synchrotron X-ray diffraction patterns with Rietveld fitting of $Ca_2Co_2O_5$ at 300, 240, 200 and 100 K. The black circle, red line, green line, blue bars and magenta line correspond to the observed data, calculated intensity, background, Bragg peaks (240 K, upper for $P2/c11$, lower for $Pcmb$; 100 K, upper for $Pcmb$, lower for $P12_1/m1$), and difference curve, respectively. Inserts show the quality of fit in $2\theta$ range of 8.882~9.012$^0$.

The 240 K pattern refined as single phase $P2/c11$ was not satisfactory (see **Fig. S4**), as it yielded relatively poor agreement factors ($R_{wp}$ =13.65%) and unacceptable visual fits to the data. The pattern was then refined as a mixture of $Pcmb$ and the $P2/c11$ phases obtained from single crystal refinement. This mixed-phase model converged to a substantially improved $R_{wp}$ = 12.87%. The phase fractions were found to be ~20 wt% $Pcmb$ and ~80 wt% $P2/c11$. This result can be interpreted to result from a sluggish first order phase transition.

The best refinement of the 100 K pattern was also obtained by a mixed-phase model ($P12_1/m1$+ $Pcmb$). The refinement converged at $R_{wp}$ =14.14% with phase fractions of ~8 wt% for $P12_1/m1$ and ~92 wt% for $Pcmb$. Refinement with single phase $Pcmb$ converged to $R_{wp}$ =15.52% with a significantly poorer visual agreement between model and data (see **Fig. S5**).

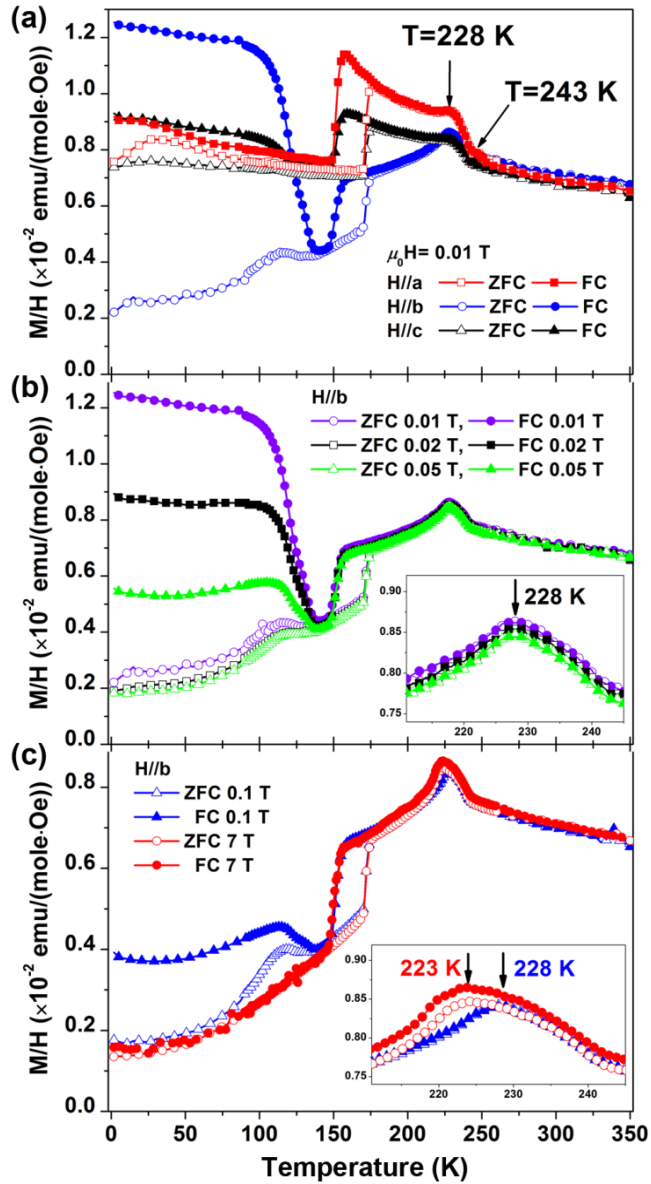

**Fig.6**. (a) Temperature dependence of the dc magnetic susceptibility of $Ca_2Co_2O_5$ in fixed field (0.01 T) for H//a (red square), H//b (blue circle), and H//c (black triangle) in ZFC (open) and FC (solid) modes. (*b-c*) The dc magnetic susceptibility versus temperature for ZFC and FC protocols at various fields along the *b* axis. Insets show details of the cusp feature around 230 K. Note the cooling and warming rates are slow (2 K/min between 350 and 260 K, 0.5 K/min between 260 and 90 K, and 2 K/min between 90 and 1.8 K).

**Magnetic Susceptibility**. The thermal evolution of the dc magnetic susceptibility for the $Ca_2Co_2O_5$ single crystal is shown in **Fig. 6(*a*)**. In the range of 243-350 K the susceptibility is isotropic with no irreversibility between zero-field cooled (ZFC) and field-cooled (FC) data, characteristic of a paramagnetic state. At *T*=243 K, the susceptibility turns higher, with anisotropy developing. We attribute this feature to long-range antiferromagnetic order, an assignment supported by heat capacity measurements (not shown) and preliminary neutron powder diffraction data.[48] With further decreasing temperature, a broad maximum at *T*=228 K is observed that shifts to 223 K in a magnetic field of 7 T (**Fig. 6(*c*), inset**), supporting the conclusion that antiferromagnetic ordering occurs within this temperature range. A hysteresis between ZFC and FC data develops between 150-180 K, a consequence of the re-entrant first order phase transition from $P12_1/m1$ to *Pcmb*.

Below 140 K, the ZFC and FC curves diverge, and the FC curve increases toward a plateau, signaling a weak ferromagnetic component directed primarily along the *b* direction of the crystal.[59] **Fig. 6(*b-c*)** show the dc magnetic susceptibility for ZFC and FC protocols at various fields applied along this axis. We see that below 140 K the dc magnetic susceptibility (FC mode) decreases with increasing of external field, joining the ZFC curves when an external field of 7 T was applied. The divergence between ZFC and FC magnetic susceptibility, combined with the magnetic field dependence, point to a ferromagnetic component below 140 K whose coercivity lies between 0.1 and 7 T.

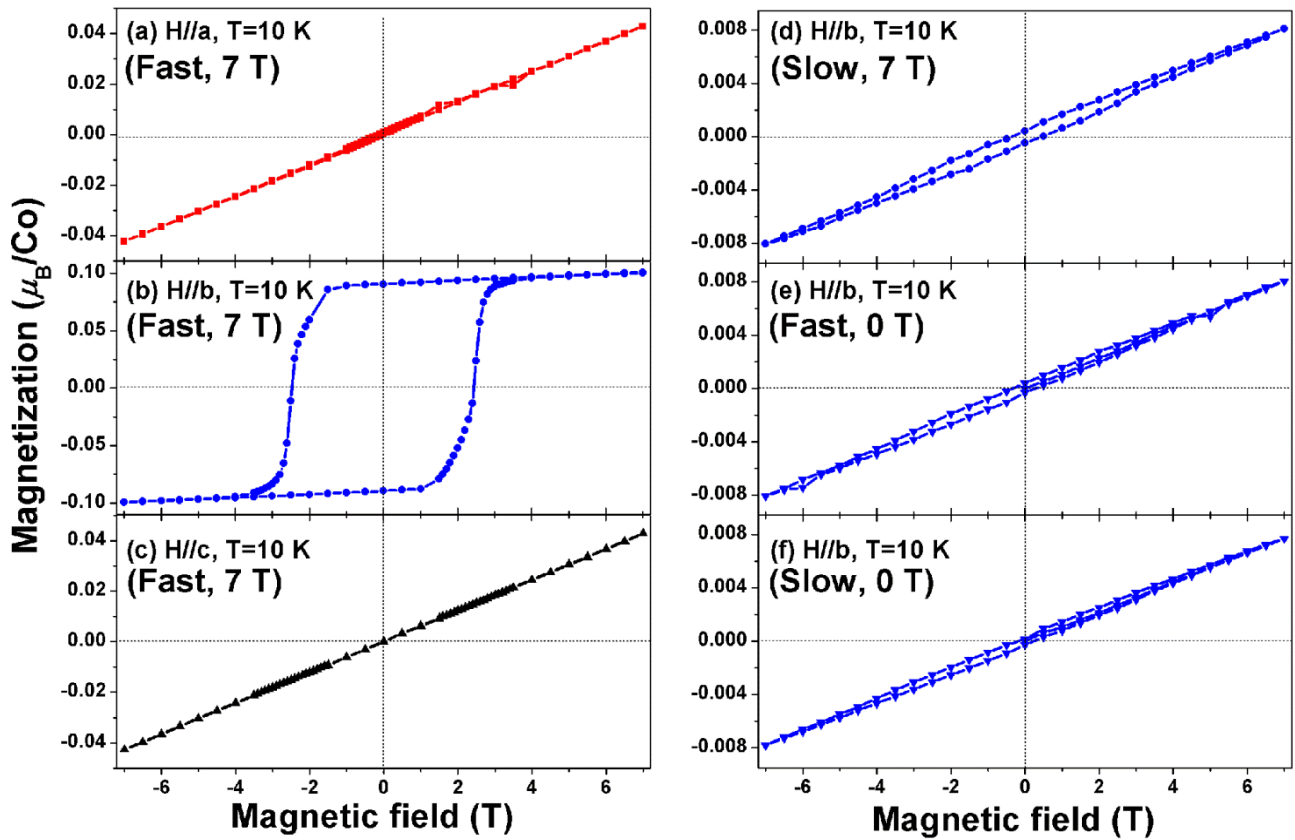

**Fig.7**. (*a-c*) Isothermal field-dependent magnetization at 10 K for *H//a*, *H//b*, and *H//c*, respectively, following fast cooling under 7 T. (*d-f*) Isothermal field-dependent magnetization at 10 K for *H//b* following different cooling rate under various fields. Fast means 10 K/min from above room temperature to 10 K; slow means 2 K/min from above room temperature to 260 K, 0.5 K/min from 260 to 90 K, and 2 K/min from 90 to 10 K. The data shown (*d*) have a small positive offset (~0.001 $\mu_B$/Co) subtracted. Details of this subtraction process can be found in the Supplemental material, **Fig. S6**. A similar offset appears in (*a&b*) but is not easily discerned at this scale and was not subtracted. We attribute this offset to a small, irreversible ferromagnetic component not arising from $Ca_2Co_2O_5$.

**Figs. 7 (*a-c*)** present the isothermal magnetization data measured at 10 K following fast (10 K/min) cooling under a field of 7 T. A pronounced hysteresis with a large coercivity (~2.5 T) is found for field directed along the *b* direction alone, confirming the existence of weak ferromagnetism in the specimen. The magnetic response of the $Ca_2Co_2O_5$ specimen depends markedly on the cooling rate and field history, **Figs. 7(*b, d-f*)**. Specifically, under ZFC, the 10 K magnetic state is insensitive to the cooling protocol, and a miniscule ferromagnetic hysteresis is found. FC (7 T) *slowly* leads to the same end (as a consistency check, the data of **Fig. 6(*c*)** were collected in a slow-cooling FC mode, and the final state agrees quantitatively with the initial condition of **Fig. 7(*d*)**). Remarkably, changing to a fast-cooling protocol in an external field of 7 T results in a magnetization (*M*~0.1 $\mu_B$/Co, see **Fig. 7 (*b*)**) more than an order of magnitude larger than that observed using the slow cooling rate (*M*~0.008 $\mu_B$/Co, see **Fig. 6(*c*)**). These contrasting data indicate a field-assisted transition into a markedly different, hard ferromagnetic state, that has been kinetically trapped from higher temperature.

We note that the remanent moment under all of these measuring conditions is quite small, 0.0004 - 0.1 $\mu_B$/Co, and we consider two explanations for this signal: intrinsic weak ferromagnetism from spin-canting or an unidentified ferromagnetic impurity phase. In the case of spin canting, and assuming a nominal 4 $\mu_B$ moment on each $Co^{3+}$, the ~0.1 $\mu_B$/Co component along *b* implies an upper bound on the cant angle of ~2°. Small concentrations of isolated cubic perovskite inclusions, known to be ferromagnetic, have been implicated as the source the weak ferromagnetic signal found in $Sr_2Co_2O_5$,[60] and the same may be occurring in the case of the measured $Ca_2Co_2O_5$ single crystal. If so, the magnetic data of **Fig. 7** indicate that the concentration of this phase is of order 3%, assuming S=3/2 intermediate-spin $Co^{4+}$ ions (~10% for low-spin S=1/2). It is likely that we would have seen an impurity at this level in the high-resolution synchrotron X-ray diffraction data if it were crystalline, but it could have conceivably escaped our detection if it were amorphous or the particle size were very small. A more restrictive constraint is that the magnetic easy axis of this potential impurity phase must be aligned with the crystallographic *b*-axis of the main $Ca_2Co_2O_5$ structure, as the FM signal is highly anisotropic below 140 K (**Fig. 7(*a-c*)**). A possible mechanism to achieve this alignment would involve oxygen diffusing into the $Ca_2Co_2O_5$ crystal during the growth or cooling processes, filling the O vacancies in the tetrahedral layers to form nanoscale, uncorrelated regions of a putative ferromagnetic $CaCoO_{3-x}$ that would have likely escaped detection by X-ray diffraction. This mechanism would naturally lead to an epitaxial relationship with the matrix phase, and one could imagine the possibility of an alignment of the magnetic easy axis with that of the brownmillerite. Unfortunately, at this time our data do not allow us to distinguish conclusively between the intrinsic and extrinsic models for the weak ferromagnetism. Attempts to oxygenate $Ca_2Co_2O_5$ are underway. If successful, this will allow us to differentiate between these two scenarios and to explore reliably the nature of the field-induced, kinetically trapped state shown in **Fig. 7**.

**CONCLUSIONS**

Bulk brownmillerite $Ca_2Co_2O_5$ has been synthesized using a high-pressure floating zone furnace, and single crystals with dimensions up to 1.4×0.8×0.5 mm³ were obtained via optimizing growth conditions. At room temperature, $Ca_2Co_2O_5$ joins $Ca_2FeCoO_5$ and $La_{2-x}A_xMn_2O_5$ (A=Ca or Sr, *x*=0.7 and 0.8) as among the few examples of brownmillerite oxides crystallizing in the *Pcmb* space group. With decreasing temperature, $Ca_2Co_2O_5$ undergoes re-entrant phase transition series *Pcmb*→ *P*2/*c*11→ *P*12$_1$/*m*1→ *Pcmb*, unique among brownmillerites. Temperature-dependent magnetic susceptibility data reveal that $Ca_2Co_2O_5$ is antiferromagnetic below ~240 K. Magnetization data also imply the potential of a distinct, field-induced phase arising uniquely from the *P*12$_1$/*m*1 structure, revealed as kinetically trapped by a rapid-cooling protocol.

Finally, the present study demonstrates a robust platform for the synthesis of new ordered anion-deficient Co-based perovskites using high pressure oxygen gas in crystal growth. We also note that anion-deficient perovskites provide a route to other new materials and properties. For instance, Youwen et al.[61] prepared cubic $SrCoO_3$ single crystals from brownmillerite $Sr_2Co_2O_5$ by high-pressure (6.5 GPa) and high temperature (1023 K) annealing. The same may be possible with $Ca_2Co_2O_5$. More generally, the successful synthesis of bulk brownmillerite $Ca_2Co_2O_5$ provides a first glimpse at the opportunity space for new, metastable perovskites and other oxides opened by high pressure optical image growth.

## ASSOCIATED CONTENT

**Supporting Information.**

Crystallographic information (CIF files) for $Ca_2Co_2O_5$ at 300 K, 240 K, 200 K and 100 K. Dipole moment calculation (EXCEL file) for some brownmillerite oxides. Atomic coordinates and equivalent isotropic displacement parameters of $Ca_2Co_2O_5$ at various temperatures; Co-O polyhedral distortion in $Ca_2Co_2O_5$, $Ca_2FeCoO_5$, $Ca_2Co_{1.54}Ga_{0.46}O_5$, and $Sr_2Co_2O_5$; dipole moment calculation details and result; Rietveld refinement results for 300, 240, 200 and 100 K data; powder X-ray diffraction patterns of precursors; comparison of the powder X-ray diffraction pattern of $Ca_2Co_2O_5$ and $Ca_2Co_{1.54}Ga_{0.46}O_5$; high pressure product at different $pO_2$; high-resolution synchrotron X-ray pattern at 240 K with Rietveld fitting using $P2/c11$ single phase and pattern at 100 K with Rietveld fitting using $Pcmb$ single phase; isothermal field-dependent magnetization at 10 K for $H//b$ following slow cooling under 7 T and -7 T. This material is available free of charge via the Internet at http://pubs.acs.org.

## AUTHOR INFORMATION

**Corresponding Author**

*Email: junjie@anl.gov; junjie.zhang.sdu@gmail.com

**Notes**

The authors declare no competing financial interest.

## ACKNOWLEDGMENTS

This work was supported by the US Department of Energy, Office of Science, Basic Energy Sciences, Materials Science and Engineering Division. Synchrotron X-ray diffraction was carried out at the Advanced Photon Source, which is supported by the U. S. Department of Energy, Office of Science, Basic Energy Sciences. T.-H.H. acknowledges the support by the Grainger Fellowship from the Department of Physics, University of Chicago. The authors thank Dr. John A. Schlueter for assistance in collecting single-crystal X-ray diffraction data, Dr. John A. Schlueter and and Dr. Ulrich Welp for help with magnetic susceptibility data, Saul H. Lapidus, Cun Yu, and Longlong Fan for help with the high resolution powder diffraction experiment, Dr. Xiaomin Lin for help with the low temperature DSC measurements, Dr. Ashfia Huq for help with neutron powder diffraction experiment, and Dr. Fei Han for helpful discussion.

## REFERENCES

(1) Raveau, B.; Seikh, M. M. *Cobalt Oxides: From Crystal Chemistry to Physics*; Wiley-VCH Verlag & Co. KGaA: Weinheim, 2012.

(2) Limelette, P.; Hébert, S.; Hardy, V.; Frésard, R.; Simon, C.; Maignan, A. *Phys. Rev. Lett.* **2006**, *97*, 046601.

(3)  Ohta, H.; Sugiura, K.; Koumoto, K. *Inorg. Chem.* **2008**, *47*, 8429.

(4)  Barilo, S. N.; Shiryaev, S. V.; Bychkov, G. L.; Shestak, A. S.; Zhou, Z. X.; Hinkov, V.; Plakhty, V. P.; Chernenkov, Y. P.; Gavrilov, S. V.; Baran, M.; Szymczak, R.; Sheptyakov, D.; Szymczak, H. *Rev. Adv. Mater. Sci.* **2006**, *12*, 33.

(5)  Takada, K.; Sakurai, H.; Takayama-Muromachi, E.; Izumi, F.; Dilanian, R. A.; Sasaki, T. *Nature* **2003**, *422*, 53.

(6)  Caignaert, V.; Maignan, A.; Singh, K.; Simon, C.; Pralong, V.; Raveau, B.; Mitchell, J. F.; Zheng, H.; Huq, A.; Chapon, L. C. *Phys. Rev. B* **2013**, *88*, 174403.

(7)  Woermann, E.; Muan, A. *J. Inorg. Nucl. Chem.* **1970**, *32*, 1455.

(8)  Kudasov, Y. B. *Phys. Rev. Lett.* **2006**, *96*, 027212.

(9)  Agrestini, S.; Chapon, L. C.; Daoud-Aladine, A.; Schefer, J.; Gukasov, A.; Mazzoli, C.; Lees, M. R.; Petrenko, O. A. *Phys. Rev. Lett.* **2008**, *101*, 097207.

(10) Korshunov, A.; Kudasov, Y.; Maslov, D.; Pavlov, V. *Solid State Phenomen* **2009**, *152-153*, 225.

(11) Fleck, C. L.; Lees, M. R.; Agrestini, S.; McIntyre, G. J.; Petrenko, O. A. *EPL* **2010**, *90*, 67006.

(12) Agrestini, S.; Fleck, C. L.; Chapon, L. C.; Mazzoli, C.; Bombardi, A.; Lees, M. R.; Petrenko, O. A. *Phys. Rev. Lett.* **2011**, *106*, 197204.

(13) Kamiya, Y.; Batista, C. D. *Phys. Rev. Lett.* **2012**, *109*, 067204.

(14) Li, S.; Funahashi, R.; Matsubara, I.; Ueno, K.; Yamada, H. *J. Mater. Chem.* **1999**, *9*, 1659.

(15) Li, S.; Funahashi, R.; Matsubara, I.; Ueno, K.; Sodeoka, S.; Yamada, H. *Chem. Mater.* **2000**, *12*, 2424.

(16) Maignan, A.; Hébert, S.; Pelloquin, D.; Michel, C.; Hejtmanek, J. *J. Appl. Phys.* **2002**, *92*, 1964.

(17) Funahashi, R.; Urata, S.; Sano, T.; Kitawaki, M. *J. Mater. Res.* **2003**, *18*, 1646.

(18) Liu, Y.; Lin, Y.; Shi, Z.; Nan, C.-W.; Shen, Z. *J. Am. Ceram. Soc.* **2005**, *88*, 1337.

(19) Wang, X. L.; Sakurai, H.; Takayama-Muromachi, E. *J. Appl. Phys.* **2005**, *97*, 10M519.

(20) Vidyasagar, K.; Gopalakrishnan, J.; Rao, C. N. R. *Inorg. Chem.* **1984**, *23*, 1206.

(21) Sharma, N.; Shaju, K. M.; Subba Rao, G. V.; Chowdari, B. V. R. *Electrochim. Acta* **2004**, *49*, 1035.

(22) Funahashi, R.; Matsubara, I.; Ikuta, H.; Takeuchi, T.; Mizutani, U.; Sodeoka, S. *Jpn. J. Appl. Phys., Part 2* **2000**, *39*, L1127.

(23) Zhang, Y.; Zhang, J.; Lu, Q. *J. Alloys Compd.* **2005**, *399*, 64.

(24) Lan, J.; Lin, Y.-H.; Li, G.-j.; Xu, S.; Liu, Y.; Nan, C.-W.; Zhao, S.-J. *Appl. Phys. Lett.* **2010**, *96*, 192104.

(25) Liu, Y.-C.; Lin, Y.-H.; Nan, C.-W.; Zhan, B.; Lan, J. *Funct. Mater. Lett.* **2013**, *06*, 1340001.

(26) Kim, M. G.; Im, Y. S.; Oh, E. J.; Kim, K. H.; Yo, C. H. *Physica B* **1997**, *229*, 338.

(27) Yong Sung, I.; Kwang Hyun, R.; Keu Hong, K.; Chul Hyun, Y. *J. Phys. Chem. Solids* **1997**, *58*, 2079.

(28) Ramezanipour, F.; Greedan, J. E.; Grosvenor, A. P.; Britten, J. F.; Cranswick, L. M. D.; Garlea, V. O. *Chem. Mater.* **2010**, *22*, 6008.

(29) Parsons, T. G.; D'Hondt, H.; Hadermann, J.; Hayward, M. A. *Chem. Mater.* **2009**, *21*, 5527.

(30) Artem, M. A.; Marina, G. R.; Evgenii, V. A. *Russ. Chem. Rev.* **2004**, *73*, 847.

(31) Lambert, S.; Leligny, H.; Grebille, D.; Pelloquin, D.; Raveau, B. *Chem. Mater.* **2002**, *14*, 1818.
(32) Istomin, S. Y.; Abdyusheva, S. V.; Svensson, G.; Antipov, E. V. *J. Solid State Chem.* **2004**, *177*, 4251.
(33) Boullay, P.; Dorcet, V.; Pérez, O.; Grygiel, C.; Prellier, W.; Mercey, B.; Hervieu, M. *Phys. Rev. B* **2009**, *79*, 184108.
(34) Steele, B. C. H. *Mater. Sci. Eng., B* **1992**, *13*, 79.
(35) Zhang, K.; Sunarso, J.; Shao, Z.; Zhou, W.; Sun, C.; Wang, S.; Liu, S. *RSC Adv.* **2011**, *1*, 1661.
(36) Abakumov, A. M.; Kalyuzhnaya, A. S.; Rozova, M. G.; Antipov, E. V.; Hadermann, J.; Van Tendeloo, G. *Solid State Sci.* **2005**, *7*, 801.
(37) Hadermann, J.; Abakumov, A. M.; D'Hondt, H.; Kalyuzhnaya, A. S.; Rozova, M. G.; Markina, M. M.; Mikheev, M. G.; Tristan, N.; Klingeler, R.; Buchner, B.; Antipov, E. V. *J. Mater. Chem.* **2007**, *17*, 692.
(38) Bruker APEX2; Bruker AXS, Inc.: Madison, Wisconsin, USA., 2005.
(39) Sheldrick, G. M.; SHELXTL version 6.12; Bruker AXS, Inc.: Madison, WI, 2001.
(40) Brown, I. D.; Altermatt, D. *Acta Crystallogr. Sect. B: Struct. Sci.* **1985**, *41*, 244.
(41) Brese, N. E.; O'Keeffe, M. *Acta Crystallogr. Sect. B: Struct. Sci.* **1991**, *47*, 192.
(42) Rodríguez-Carvajal, J.; Hennion, M.; Moussa, F.; Moudden, A. H.; Pinsard, L.; Revcolevschi, A. *Phys. Rev. B* **1998**, *57*, R3189.
(43) Larson, A. C.; Dreele, R. B. V. *General Structure Analysis System (GSAS)*, Los Alamos National Laboratory Report LAUR 86-748, 2000.
(44) Toby, B. H. *J. Appl. Crystallogr.* **2001**, *34*, 210.
(45) Stephens, P. W. *J. Appl. Crystallogr.* **1999**, *32*, 281.
(46) de la Calle, C.; Aguadero, A.; Alonso, J. A.; Fernández-Díaz, M. T. *Solid State Sci.* **2008**, *10*, 1924.
(47) Abakumov et al. (Chem. Mater. **2008**, 20, 7188) have reported a local symmetry of *P*2/*c* in the commensurate $Sr_2Fe_2O_5$ brownmillerite *via* TEM; however, there is no evidence for long-range order in this space group.
(48) Neutron powder diffraction data were collected at the Spallation Neutron Source (POWGEN) at Oak Ridge National Laboratory. The pattern collected at 300 K only shows structural peaks, while those at 200 and 130 K show additional AFM superlattice reflections. However, the AFM superlattice peaks at 200 and 130 K are totally different.
(49) Singh, S. *Phase Transitions* **2000**, *72*, 183.
(50) Suzuki, E.; Shiozaki, Y. *Phys. Rev. B* **1996**, *53*, 5217.
(51) Martin, T. D. *J. Phys.: Condens. Matter* **2011**, *23*, 225402.
(52) Leu, L.-C.; Bian, J.; Gout, D.; Letourneau, S.; Ubic, R. *RSC Adv.* **2012**, *2*, 1598.
(53) Brinkmann, C.; Eckert, H.; Wilmer, D.; Vogel, M.; auf der Günne, J. S.; Hoffbauer, W.; Rau, F.; Pfitzner, A. *Solid State Sci.* **2004**, *6*, 1077.
(54) David, R.; Pautrat, A.; Filimonov, D.; Kabbour, H.; Vezin, H.; Whangbo, M.-H.; Mentré, O. *J. Am. Chem. Soc.* **2013**, *135*, 13023.
(55) Downie, L. J.; Thompson, S. P.; Tang, C. C.; Parsons, S.; Lightfoot, P. *CrystEngComm* **2013**, *15*, 7426.
(56) Rae, A. I. M.; Dove, M. T. *J. Phys. C: Solid State Phys.* **1983**, *16*, 3233.
(57) Sullivan, E.; Hadermann, J.; Greaves, C. *J. Solid State Chem.* **2011**, *184*, 649.

(58) Ramezanipour, F.; Cowie, B.; Derakhshan, S.; Greedan, J. E.; Cranswick, L. M. D. *J. Solid State Chem.* **2009**, *182*, 153.

(59) An extremely small hysteresis can be observed in our data for *H*//*a*. While this may arise from a slight misalignment of the crystal, we cannot rule out the possibility that there is a small *a*-component to the canted moment.

(60) Muñoz, A.; de la Calle, C.; Alonso, J. A.; Botta, P. M.; Pardo, V.; Baldomir, D.; Rivas, J. *Phys. Rev. B* **2008**, *78*, 054404.

(61) Youwen, L.; Yoshio, K.; Shintaro, I.; Yasujiro, T.; Yoshinori, T. *J. Phys.: Condens. Matter* **2011**, *23*, 245601.

**Table of Contents**

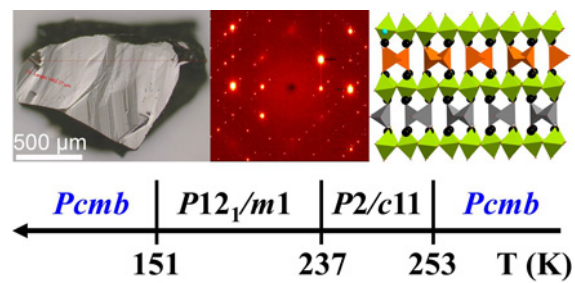

**Supporting Information for:**

# Brownmillerite $Ca_2Co_2O_5$: Synthesis, Stability, and Re-entrant Single-Crystal-to-Single-Crystal Structural Transitions

Junjie Zhang,[*,†] Hong Zheng,[†] Christos D. Malliakas,[†] Jared M. Allred,[†] Yang Ren,[§] Qing'an Li,[†] Tian-Heng Han[†,‡] and J.F. Mitchell[†]

[†]Materials Science Division, Argonne National Laboratory, Argonne, Illinois 60439, USA
[§]X-ray Science Division, Argonne National Laboratory, Argonne, Illinois, 60439, USA
[‡]The James Franck Institute and Department of Physics, University of Chicago, Chicago, IL 60637, USA

**Table S1.** Atomic coordinates ($\times 10^4$) and equivalent isotropic displacement parameters ($Å^2 \times 10^3$) of $Ca_2Co_2O_5$ at various temperatures with estimated standard deviations in parentheses.

**Table S2.** Comparison of cobalt polyhedral distortion in $Ca_2Co_2O_5$, $Ca_2FeCoO_5$, $Ca_2Co_{1.54}Ga_{0.46}O_5$, and $Sr_2Co_2O_5$.

**Table S3.** Summary of space groups, interlayer separation, degree of twisting of the tetrahedral chains and the dipole moment of some brownmillerite oxides.

**Table S4.** Parameters from the final Rietveld refinement at each temperature for $Ca_2Co_2O_5$.

**Fig. S1.** Powder X-ray diffraction patterns of precursors.

**Fig. S2.** Comparison of the powder X-ray diffraction pattern of $Ca_2Co_2O_5$ and $Ca_2Co_{1.54}Ga_{0.46}O_5$.

**Fig. S3.** High pressure product at different $pO_2$.

**Fig. S4.** (a) High-resolution synchrotron X-ray pattern at 240 K with Rietveld fitting using $P2/c11$ single phase; (b) Zoom around 8.95º; (c) zoom around 6.42º.

**Fig. S5.** (a) High-resolution synchrotron X-ray pattern at 100 K with Rietveld fitting using $Pcmb$ single phase; (b) Zoom around 8.95º; (c) zoom around 12.45º.

**Fig. S6.** (a) Isothermal field-dependent magnetization at 10 K for $H//b$ following slow cooling (2 K/min from above room temperature to 260 K, 0.5 K/min from 260 to 90 K, then 2 K/min from 90 to 10 K) under 7 T and -7 T ($M_{+7T}$ and $M_{-7T}$, respectively) showing vertical offset whose sign depends on that of the cooling field. (b) The function ($M_{+7T}$-$M_{-7T}$)/2, which removes the offset observed in (a).

**Table S1.** Atomic coordinates (×10⁻⁴) and equivalent isotropic displacement parameters ($\text{Å}^2\times 10^{-3}$) of $Ca_2Co_2O_5$ from single crystal X-ray diffraction data at various temperatures.

| Label | x | y | z | Occupancy | $U_{eq}^*$ | Label | x | y | z | Occupancy | $U_{eq}^*$ |
|---|---|---|---|---|---|---|---|---|---|---|---|
| **300 K** | | | | | | **100 K** | | | | | |
| Ca(1) | -125(1) | 1086(1) | 7597(2) | 1 | 10(1) | Ca(1) | -200(1) | 1078(1) | 2400(1) | 1 | 5(1) |
| Ca(2) | 4908(1) | 3915(1) | 4904(2) | 1 | 10(1) | Ca(2) | 4892(1) | 3922(1) | 5099(1) | 1 | 6(1) |
| Co(1) | 4509(1) | 2500 | 7200(1) | 1 | 9(1) | Co(1) | 4462(1) | 2500 | 2804(1) | 1 | 5(1) |
| Co(2) | -458(1) | 2500 | 5298(1) | 1 | 9(1) | Co(2) | -468(1) | 2500 | 4702(1) | 1 | 5(1) |
| Co(3) | 0 | 0 | 5000 | 1 | 7(1) | Co(3) | 0 | 0 | 5000 | 1 | 4(1) |
| Co(4) | -5047(1) | 0 | 7500 | 1 | 7(1) | Co(4) | -5107(1) | 0 | 2500 | 1 | 3(1) |
| O(1) | 896(4) | 2500 | 6927(2) | 1 | 11(1) | O(1) | 879(4) | 2500 | 3084(2) | 1 | 6(1) |
| O(2) | 5928(4) | 2500 | 5576(2) | 1 | 11(1) | O(2) | 5954(4) | 2500 | 4400(2) | 1 | 6(1) |
| O(3) | 190(3) | 3602(3) | 4704(5) | 1 | 13(1) | O(3) | 223(3) | 3608(2) | 5307(3) | 1 | 9(1) |
| O(4) | 5088(3) | 1396(3) | 7796(5) | 1 | 15(1) | O(4) | 5045(3) | 1386(2) | 2198(3) | 1 | 9(1) |
| O(5) | -2440(4) | -99(2) | 3742(5) | 1 | 11(1) | O(5) | -2367(3) | -102(2) | 6257(2) | 1 | 6(1) |
| O(6) | -2566(4) | 140(2) | 6241(5) | 1 | 10(1) | O(6) | -2631(3) | 150(2) | 3762(2) | 1 | 6(1) |
| **240 K** | | | | | | **200 K** | | | | | |
| Ca(1) | 9808(2) | 6074(1) | 7405(2) | 1 | 9(1) | Ca(1) | 205(2) | 1079(1) | 2604(1) | 1 | 8(1) |
| Ca(2) | 148(2) | 1082(1) | 2603(2) | 1 | 9(1) | Ca(2) | 202(2) | 6078(1) | 2405(1) | 1 | 8(1) |
| Ca(3) | 4903(2) | 1081(1) | 4905(2) | 1 | 9(1) | Ca(3) | 4900(2) | 6081(1) | 102(1) | 1 | 8(1) |
| Ca(4) | 5095(2) | 6077(1) | 5102(2) | 1 | 9(1) | Ca(4) | 4898(2) | 1078(1) | 4904(1) | 1 | 8(1) |
| Co(1) | 5525(1) | 2512(1) | 2193(1) | 1 | 8(1) | Co(1) | 4454(2) | 2500 | 7191(1) | 1 | 7(1) |
| Co(2) | 466(1) | 2494(1) | 300(1) | 1 | 8(1) | Co(2) | 5548(2) | 2500 | 2195(1) | 1 | 8(1) |
| Co(3) | 0 | 0 | 0 | 1 | 6(1) | Co(3) | 462(2) | 2500 | 297(1) | 1 | 8(1) |
| Co(4) | 0 | 5000 | 0 | 1 | 7(1) | Co(4) | 9528(2) | 2500 | 5299(1) | 1 | 7(1) |
| Co(5) | 5067(1) | 0 | 2500 | 1 | 6(1) | Co(5) | 0 | 0 | 5000 | 1 | 6(1) |
| Co(6) | 4898(2) | 5000 | 7500 | 1 | 5(1) | Co(6) | 0 | 0 | 0 | 1 | 6(1) |
| | | | | | | Co(7) | 5112(1) | 1(1) | 2503(1) | 1 | 6(1) |
| O(1) | 4061(5) | 2506(2) | 590(3) | 1 | 10(1) | O(1) | 878(5) | 2500 | 6924(3) | 1 | 8(1) |
| O(2) | -877(4) | 2500(2) | 1923(2) | 1 | 9(1) | O(2) | 9132(5) | 2500 | 1927(3) | 1 | 9(1) |
| O(3) | 4924(4) | 1417(3) | 2800(4) | 1 | 12(1) | O(3) | 5942(5) | 2500 | 5597(3) | 1 | 8(1) |
| O(4) | 4978(4) | 3644(3) | 2780(4) | 1 | 13(1) | O(4) | 4056(5) | 2500 | 593(3) | 1 | 9(1) |
| O(5) | -228(5) | 3589(3) | -310(4) | 1 | 13(1) | O(5) | 217(5) | 1394(2) | 4703(3) | 1 | 12(1) |
| O(6) | 2384(5) | 5097(2) | -1259(3) | 1 | 9(1) | O(6) | 234(5) | 6402(2) | 308(3) | 1 | 11(1) |
| O(7) | 7370(5) | 5143(2) | 8755(3) | 1 | 9(1) | O(7) | 4971(5) | 6382(2) | 2206(3) | 1 | 12(1) |
| O(8) | 2417(5) | -102(2) | -1258(3) | 1 | 9(1) | O(8) | 4963(5) | 1386(2) | 2806(3) | 1 | 13(1) |
| O(9) | 2584(5) | 146(2) | 1244(3) | 1 | 9(1) | O(9) | 2382(4) | 98(2) | 6263(3) | 1 | 9(1) |
| O(10) | -189(5) | 1373(3) | -293(4) | 1 | 12(1) | O(10) | 7639(4) | 103(2) | 1255(3) | 1 | 9(1) |
| | | | | | | O(11) | 2620(4) | 5146(2) | 3765(3) | 1 | 9(1) |
| | | | | | | O(12) | 2664(4) | 149(2) | 1255(3) | 1 | 8(1) |

*$U_{eq}$ is defined as one third of the trace of the orthogonalized $U_{ij}$ tensor.

**Table S2.** Comparison of cobalt polyhedral distortion in $Ca_2Co_2O_5$, $Ca_2FeCoO_5$, $Ca_2Co_{1.54}Ga_{0.46}O_5$, and $Sr_2Co_2O_5$.

| Compound | Temperature (K) | Tetrahedral distortion ($\times 10^{-4}$) | Octahedral distortion ($\times 10^{-4}$) |
|---|---|---|---|
| $Ca_2Co_2O_5$ | 300 | 12.7, 11.9 (average: 12.3) | 23.5, 22.4 (average: 23.0) |
|  | 240 | 11.7, 10.5 (average: 11.1) | 13.6, 25.7, 24.9, 8.6 (average: 18.2) |
|  | 200 | 9.4, 10.2, 13.7, 11.5 (average: 11.2) | 19.4, 22.8, 15.1 (average: 19.1) |
|  | 100 | 11.1, 10.3 (average: 10.7) | 19.6, 13.1 (average: 16.4) |
| $Ca_2FeCoO_5$[S1] | 300 | 8.2 | 13.4 |
| $Ca_2Co_{1.54}Ga_{0.46}O_5$[S2] | 300 | 6.9 | 30.6 |
| $Sr_2Co_2O_5$[S3] | 300 | 11.9 | 45.8 |
|  | 200 | 5.7 | 45.6 |
|  | 100 | 3.5 | 44.0 |

**Table S3** Summary of space groups, interlayer separation ($b/2$), degree of twisting of the tetrahedral chains (O-O-O angle) and dipole moment for some brownmillerite oxides.

| Compound | Space group[a] | b/2 (Å) | O-O-O angle (°) | Dipole moment of each tetrahedral (Debye) | Net dipole moment along each chain | | Tetrahedral site |
|---|---|---|---|---|---|---|---|
|  |  |  |  |  | $\times 10^{-4}$ esu·cm/Å$^3$ | Debye |  |
| $Ca_2FeAlO_5$[S4] | $I2mb$ | 7.25 | 131.5 | $FeO_4$: 2.45; $AlO_4$: 2.61 | 16.8 | 1.4 | Disorder |
| $Ca_2FeGaO_5$[S5] | $Pnma$ | 7.35 | 125.8 | $GaO_4$: 1.70 | 17.7 | 1.6 | Order |
| $Ca_2Fe_{1.66}V_{0.34}O_5$[S6] | $Pnma$ | 7.37 | 125.8 | $FeO_4$: 2.46 | 20.2 | 1.8 | Order |
| $Ca_2Fe_2O_5$[S7] | $Pnma$ | 7.41 | 125.3 | $FeO_4$: 2.16 | 21.4 | 1.9 | Order |
| $Ca_2FeCoO_5$[S1] | $Pcmb$ | 7.41 | 123.5 | $CoO_4$: 2.56; $FeO_4$: 3.00 | 33.8 | 3.0 | Order |
| $Ca_2Co_{1.54}Ga_{0.46}O_5$[S2] | $Pnma$ | 7.44 | 123.7 | $CoO_4$: 2.60; $GaO_4$: 2.58 | 40.4 | 3.5 | Disorder |
| $Ca_2Co_2O_5$ | $Pcmb$ | 7.46 | 121.5 | $CoO_4$: 3.35, 3.33 | 45.2 | 3.9 | Order |
| $SrCaMnGaO_5$[S8] | $I2mb$ | 7.89 | 124.2 | $GaO_4$: 3.39 | 35.0 | 3.2 | Order |
| $LaCaCuGaO_5$[S9] | $I2mb$ | 7.92 | 124.5 | $GaO4$: 2.21 | 22.5 | 2.1 | Order |
| $La_{1.2}Sr_{0.8}Mn_2O_5$[S10] | $Pcmb$ | 8.31 | 111.8 | $MnO_4$: 5.20, 5.01 | 63.4 | 6.3 | Order |

**Calculation Details**

A bond-valence approach has been used to calculate the direction and magnitude of the dipole moments for each tetrahedron. The local dipole moment of a tetrahedron can be calculated using the Debye equation, $\mu=neR$ ($\mu$ is the net dipole moment in Debye, $n$ is the total number of electrons, $e$ is the charge on an electron, and $R$ is the distance in cm between the "centroids" of positive and negative charge).[S11,S12] The distribution of electrons on each atom was estimated using bond valence theory.[S13,S14] To calculate the net dipole moment along each chain, we normalized the unit cell to dimensions ~5×11×15 Å$^3$. For $Ca_2FeAlO_5$ and $Ca_2Co_{1.54}Ga_{0.46}O_5$ with disordered atoms in tetrahedral sites, the net dipole moment along each chain was calculated by taking into account the statistical occupancy.

**Table S4.** Parameters from the final Rietveld refinement at 300, 240, 200 and 100 K for $Ca_2Co_2O_5$.

| T (K) | Space Group | Phase (wt%) | Lattice parameters | | | | | Quality of Fit | | |
|---|---|---|---|---|---|---|---|---|---|---|
| | | | $a$ (Å) | $b$ (Å) | $c$ (Å) | $\beta$ (°) | $V$ (Å$^3$) | $R_{wp}$ | $R_p$ | $\chi^2$ |
| 300 | *Pcmb* | 100 | 5.288203(13) | 14.924582(38) | 10.953042(23) | - | 864.460(2) | 13.46% | 10.91% | 2.262 |
| 240 | *Pcmb* | 20 | 5.291507 (100) | 14.851301 (247) | 10.962284 (117) | - | 861.479(18) | 12.87% | 9.55% | 2.938 |
| | *P2/c11* | 80 | 5.292553 (27) | 14.807267 (89) | 10.964293(34) | 90.158 | 859.249(6) | | | |
| 200 | *P12$_1$/m1* | 100 | 5.299726(15) | 14.781377 (45) | 10.976010(22) | 90.329 | 859.816(3) | 12.22% | 9.03% | 2.874 |
| 100 | *P12$_1$/m1* | 8 | 5.299893(62) | 14.750240(240) | 10.982167(146) | 90.406 | 858.506(15) | 14.14% | 10.25% | 4.210 |
| | *Pcmb* | 92 | 5.302917(11) | 14.749347(35) | 10.986770(22) | - | 859.325(2) | | | |

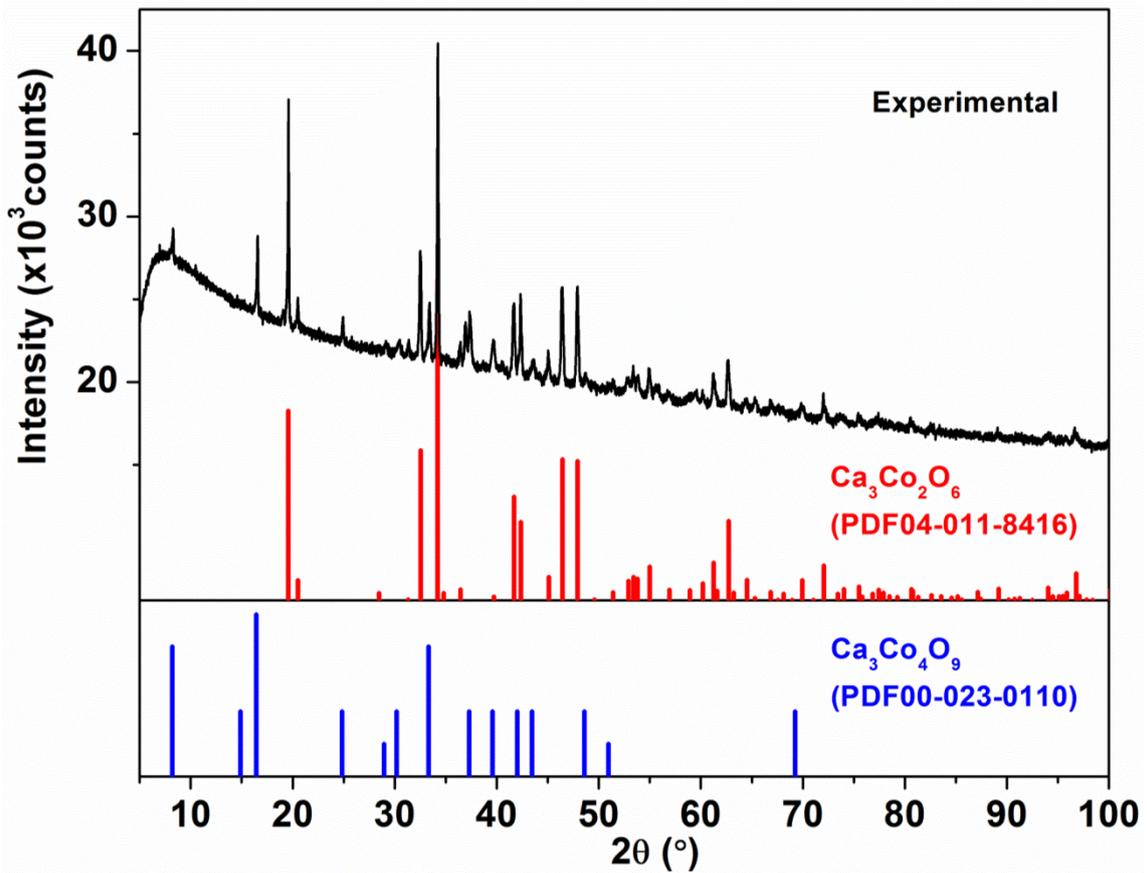

**Fig. S1.** Powder X-ray diffraction patterns of precursors.

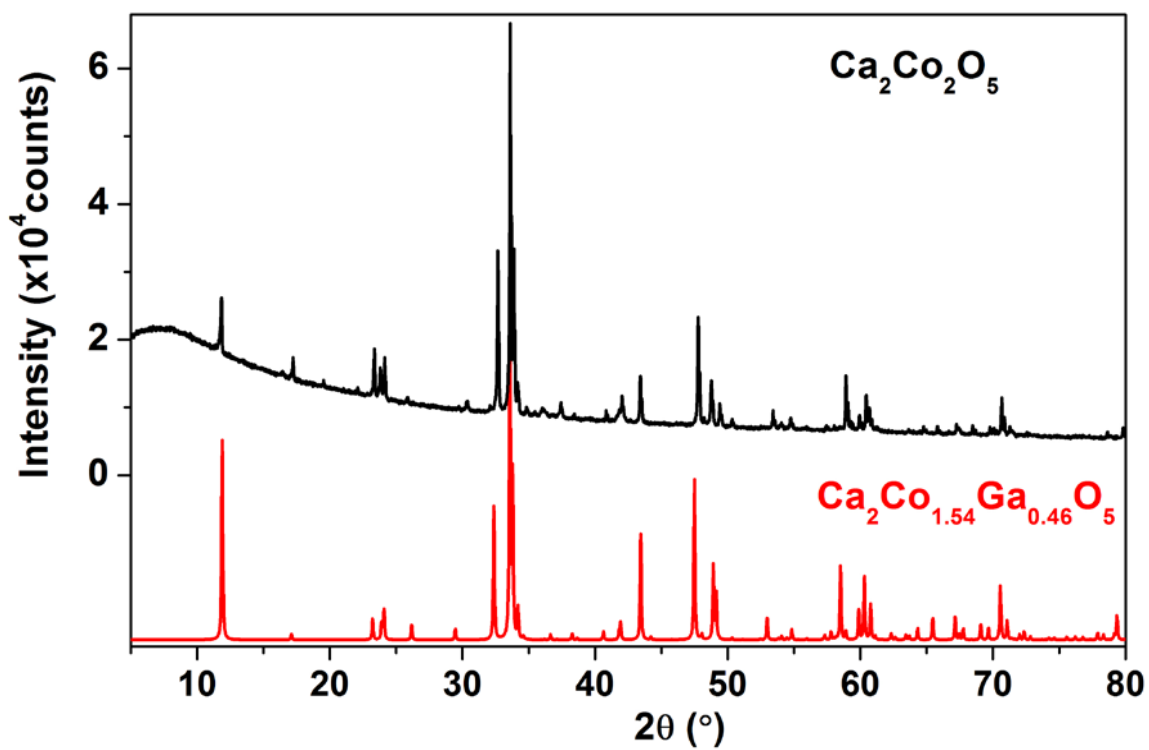

**Fig. S2.** Comparison of the powder X-ray diffraction pattern of $Ca_2Co_2O_5$ and $Ca_2Co_{1.54}Ga_{0.46}O_5$. The latter pattern is calculated from the single crystal structure in Ref. S2.

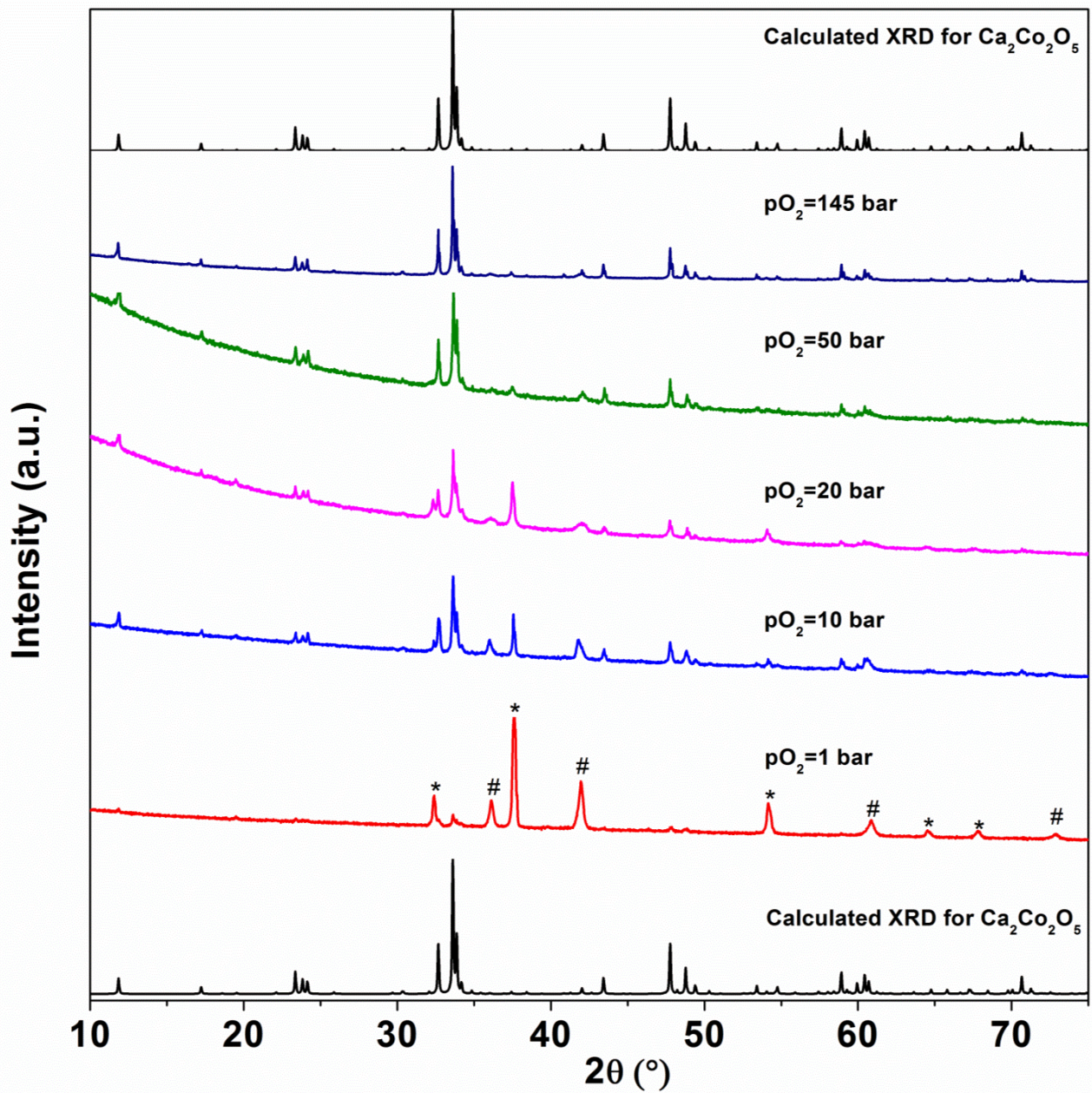

**Fig. S3.** Synthesis products as a function of pO$_2$. Note CaO (*) and CoO (#).

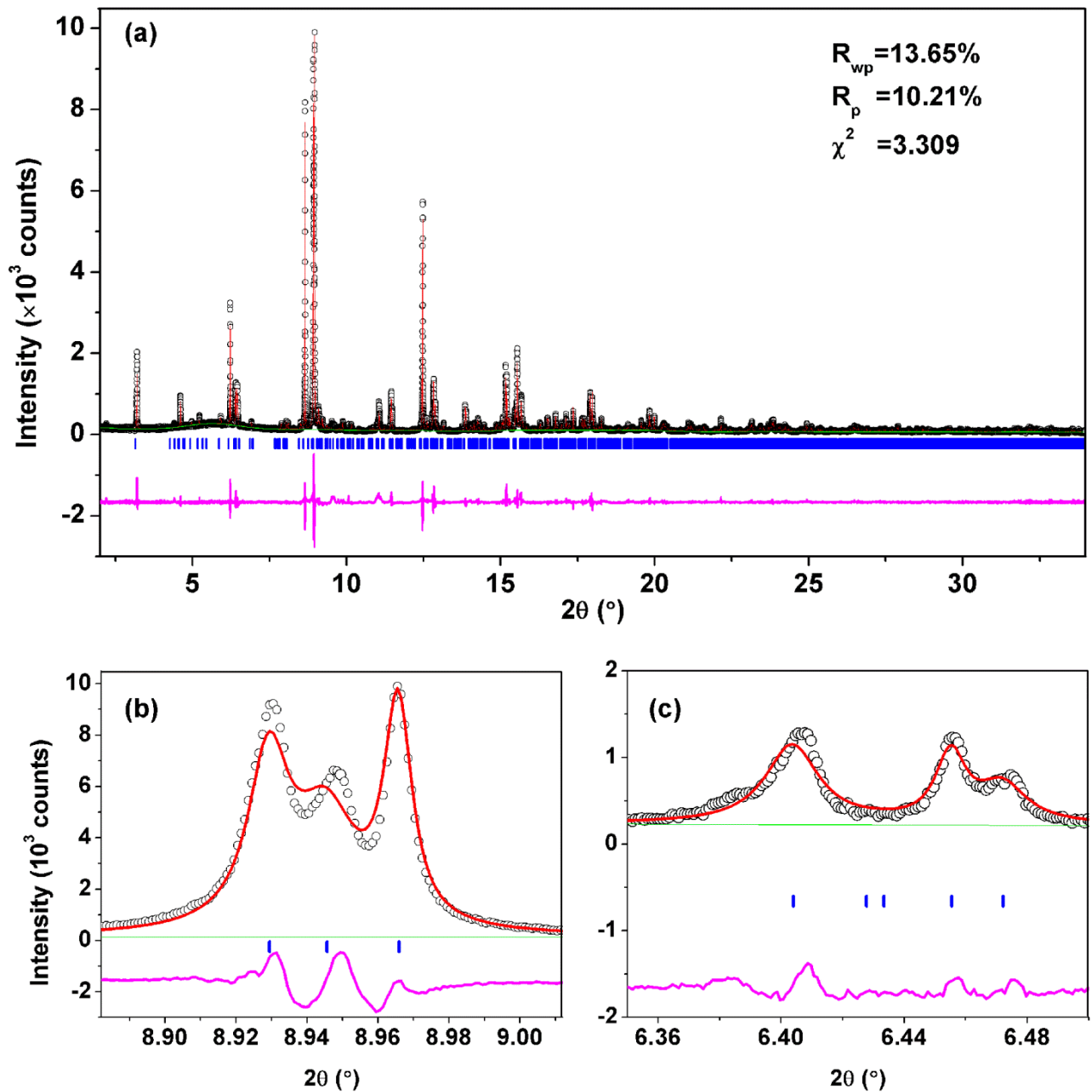

**Fig. S4.** (*a*) High-resolution synchrotron X-ray pattern at 240 K with Rietveld fitting using a single *P*2/*c*11 phase; (*b*) detail around 8.95°; (*c*) detail around 6.42°. The black circle, red line, green line, blue bars and magenta line correspond to the observed data, calculated intensity, background, Bragg peaks, and difference curve, respectively. Compared to the two-phase (*P*2/*c*11+ *Pcmb*) refinement in Fig. 5(*b*) of the main text, the fit is markedly worse.

-7-

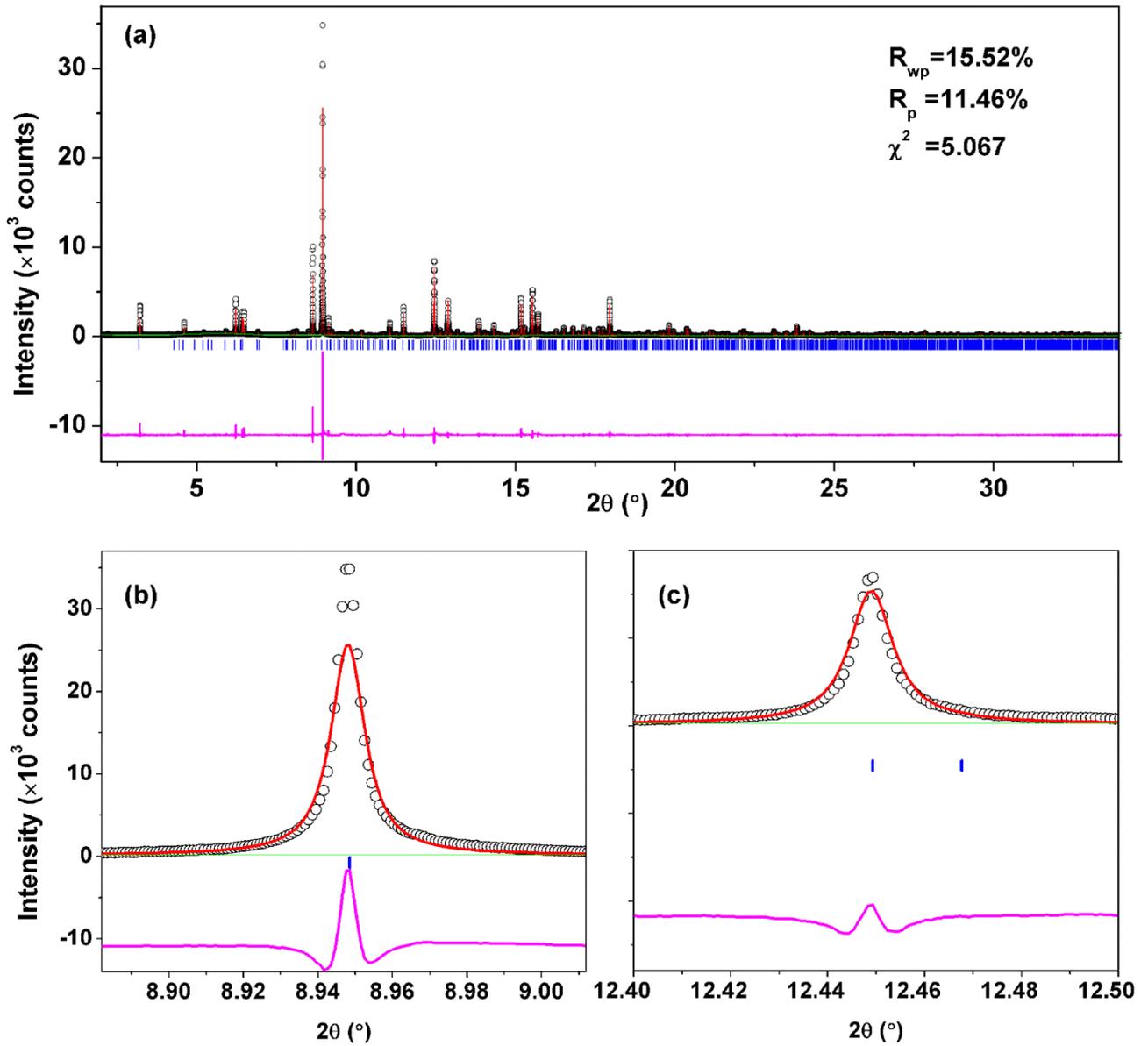

**Fig. S5.** (*a*) High-resolution synchrotron X-ray pattern at 100 K with Rietveld fitting using *Pcmb* single phase; (*b*) detail around 8.95°; (*c*) detail around 12.45°. The black circle, red line, green line, blue bars and magenta line correspond to the observed data, calculated intensity, background, Bragg peaks, and difference curve, respectively. Compared to the two-phase (*P*12$_1$/*m*1+ *Pcmb*) refinement in Fig. 5(*d*) of the main text, this fit is markedly worse.

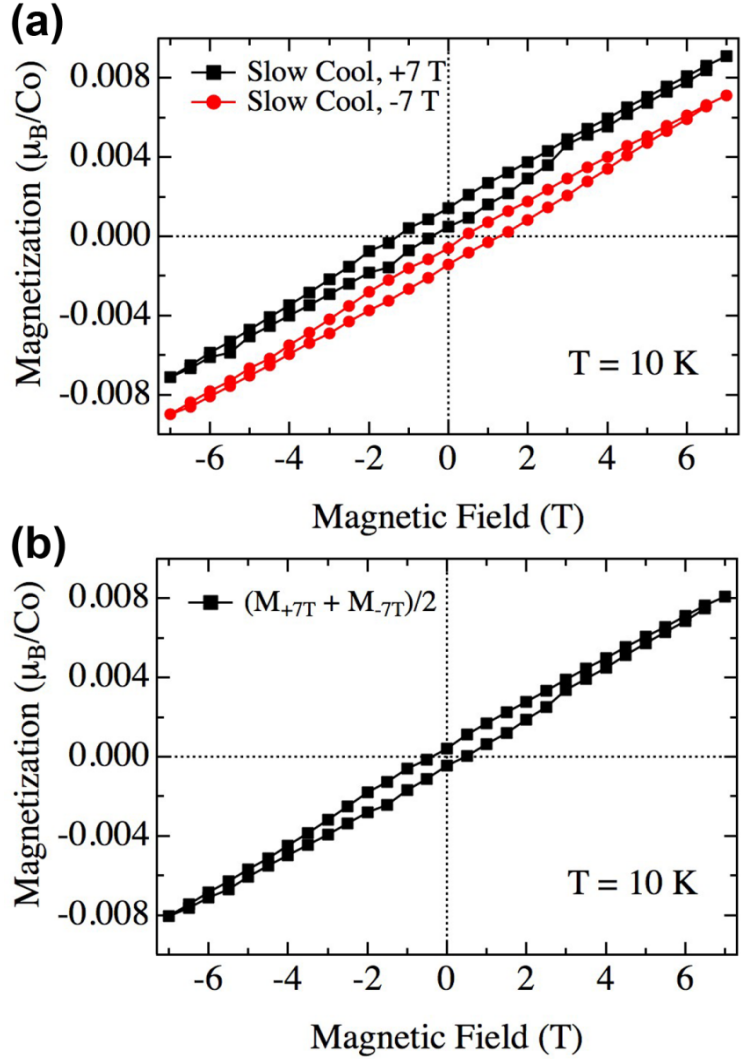

**Fig. S6.** (*a*) Isothermal field-dependent magnetization at 10 K for *H*//*b* following slow cooling (2 K/min from above room temperature to 260 K, 0.5 K/min from 260 to 90 K, then 2 K/min from 90 to 10 K) under 7 T and -7 T ($M_{+7T}$ and $M_{-7T}$, respectively) showing vertical offset whose sign depends on that of the cooling field. (*b*) The function $(M_{+7T}-M_{-7T})/2$, which removes the offset observed in (*a*).

**Fig. S6** shows the isothermal field-dependent magnetization along the *b*-axis following slow cooling (2 K/min from above room temperature to 260 K, 0.5 K/min from 260 to 90 K, and 2 K/min from 90 to 10 K) to 10 K in either +7 T or -7 T. A small vertical offset of the data is observed in either case, with opposite sign. We conclude that this small offset (not observed in the ZFC data, see **Fig. 7** of main text) arises from a component in the sample that becomes magnetized during the field-cooling process. The nature of this extremely small ferromagnetic component is not known. However, the data of **Fig. S6(*a*)** demonstrate that it cannot be reversed by application of fields as high as 7 T. The small data offset (amounting to ~0.001 $\mu_B$/Co) can be subtracted by computing $(M_{+7T}-M_{-7T})/2$, with the result shown in **Fig. S6(*b*)**, the same data plotted in Fig 7(d) of the main text. The remanent moment of ~0.0004 $\mu_B$/Co represents either the upper bound on the ordered moment of $Ca_2Co_2O_5$ under these conditions (almost three orders of magnitude less than that found for the fast cooling protocol (**Fig. 7(*b*)** main text) or a signal arising from some other unidentified weakly ferromagnetic impurity phase.

-9-

# References

(S1) Ramezanipour, F.; Greedan, J. E.; Grosvenor, A. P.; Britten, J. F.; Cranswick, L. M. D.; Garlea, V. O. *Chem. Mater.* **2010**, *22*, 6008.

(S2) Istomin, S. Y.; Abdyusheva, S. V.; Svensson, G.; Antipov, E. V. *J. Solid State Chem.* **2004**, *177*, 4251.

(S3) Sullivan, E.; Hadermann, J.; Greaves, C. *J. Solid State Chem.* **2011**, *184*, 649.

(S4) Jupe, A. C.; Cockcroft, J. K.; Barnes, P.; Colston, S. L.; Sankar, G.; Hall, C. *J. Appl. Crystallogr.* **2001**, *34*, 55.

(S5) Arpe, R.; Müller-Buschbaum, H.; Schenck, R. V. *Z. Anorg. Allg. Chem.* **1974**, *410*, 97.

(S6) Harringer, N. A.; Presslinger, H.; Klepp, K. O. *Z. Kristallogr. - New Cryst. Struct.* **2004**, *219*, 5.

(S7) Berastegui, P.; Eriksson, S. G.; Hull, S. *Mater. Res. Bull.* **1999**, *34*, 303.

(S8) Battle, P. D.; Bell, A. M. T.; Blundell, S. J.; Coldea, A. I.; Gallon, D. J.; Pratt, F. L.; Rosseinsky, M. J.; Steer, C. A. *J. Solid State Chem.* **2002**, *167*, 188.

(S9) Luzikova, A. V.; Kharlanov, A. L.; Antipov, E. V.; Müller-Buschbaum, H. *Z. Anorg. Allg. Chem.* **1994**, *620*, 326.

(S10) Parsons, T. G.; D'Hondt, H.; Hadermann, J.; Hayward, M. A. *Chem. Mater.* **2009**, *21*, 5527.

(S11) Maggard, P. A.; Nault, T. S.; Stern, C. L.; Poeppelmeier, K. R. *J. Solid State Chem.* **2003**, *175*, 27.

(S12) Zhang, J.; Zhang, Z.; Zhang, W.; Zheng, Q.; Sun, Y.; Zhang, C.; Tao, X. *Chem. Mater.* **2011**, *23*, 3752.

(S13) Brown, I. D.; Altermatt, D. *Acta Crystallogr. Sect. B: Struct. Sci.* **1985**, *41*, 244.

(S14) Brese, N. E.; O'Keeffe, M. *Acta Crystallogr. Sect. B: Struct. Sci.* **1991**, *47*, 192.